\documentclass{lmcs} 
\pdfoutput=1

\usepackage{lastpage}
\lmcsdoi{20}{2}{9}
\lmcsheading{}{\pageref{LastPage}}{}{}%
{Feb.~01,~2023}{May~17,~2024}{}


\usepackage{hyperref}

\theoremstyle{plain} 

\usepackage[utf8]{inputenc} 
\usepackage[T1]{fontenc}    

\usepackage{hyperref}       
\usepackage{url}            
\usepackage{booktabs}       
\usepackage{amsfonts}       
\usepackage{amssymb}     
\usepackage{mathtools}
\usepackage{amsmath}
\usepackage{mathrsfs}
\usepackage{nicefrac}       
\usepackage{microtype}      
\usepackage{tikz-cd}
\usepackage{enumitem}
\usepackage{comment}
\usepackage{stmaryrd}
\usepackage{float}

\newcommand{\As}{\mathcal{A}}
\newcommand{\Bs}{\mathcal{B}}
\newcommand{\cala}{\mathcal{A}}
\newcommand{\calb}{\mathcal{B}}
\newcommand{\Cs}{\mathcal{C}}

\newcommand{\Pk}{\mathbb{P}_{k}}
\newcommand{\Ek}{\mathbb{E}_{k}}
\newcommand{\Rk}{\mathbb{PR}_{k}}
\newcommand{\Rkn}{\mathbb{PR}_{k,n}}

\newcommand{\PR}{\mathbb{PR}}

\newcommand{\T}{\mathbb{T}}
\newcommand{\CC}{\mathbb{C}}

\newcommand{\Peb}{\textbf{Peb}}
\newcommand{\GG}{\normalfont\textbf{G}}
\newcommand{\PebR}{\normalfont\textbf{PebR}}
\newcommand{\PPeb}{\normalfont\textbf{PPeb}}

\newcommand{\last}{\textsf{last}}
\newcommand{\pebbles}{\textsf{pebbles}}
\newcommand{\Active}{\textsf{Active}}
\newcommand{\Rsig}{\mathfrak{R}(\sigma)}
\newcommand{\Rsigf}{\mathfrak{R}_f(\sigma)}
\newcommand{\Rsigplus}{\mathfrak{R}(\sigma^{+})}
\newcommand{\Rsigplusf}{\mathfrak{R}_f(\sigma^{+})}

\newcommand{\TsigCk}{\mathfrak{T}^{\CC_{k}}(\sigma)}

\newcommand{\TsigRk}{\mathfrak{T}^{\mathbb{PR}_{k}}(\sigma)}

\newcommand{\setn}{[\textsf{n}]}
\newcommand{\setm}{[\textsf{m}]}
\newcommand{\setk}{[\textsf{k}]}

\newcommand{\setr}{[\textsf{r}]}
\newcommand{\darr}{\downarrow}
\newcommand{\uarr}{\uparrow}

\newcommand{\emptyseq}{\epsilon}
\newcommand{\id}{\mathsf{id}}
\newcommand{\Gf}{\mathcal{G}}
\newcommand{\arc}{\frown}

\newcommand{\pw}{\mathsf{pw}}
\newcommand{\diag}{\mathbf{diag}}
\newcommand{\tp}{\mathbf{tp}}

\newcommand{\Logic}{\mathscr{J}}
\newcommand{\pLogic}{\exists^{+}\Logic}

\newcommand{\cLogic}{\#\Logic}

\newcommand{\LogicW}{\mathscr{L}_{\infty}}
\newcommand{\pLogicW}{\exists^{+}\LogicW}

\newcommand{\LogicK}{\mathscr{L}^{k}}
\newcommand{\pLogicK}{\exists^{+}\LogicK}

\newcommand{\cLogicK}{\#\LogicK}

\newcommand{\RLogicK}{{\curlywedge}\mathscr{L}^{k}}
\newcommand{\pRLogicK}{{\exists^{+}\RLogicK}}
\newcommand{\eRLogicK}{{\exists\RLogicK}}

\newcommand{\cRLogicK}{{\#\RLogicK}}

\newcommand{\equivL}{\equiv^\Logic}
\newcommand{\presL}{\Rrightarrow^\Logic}
\newcommand{\IFF}{\Leftrightarrow}
\newcommand{\Kl}{\mathcal{K}}
\newcommand{\EM}{\mathcal{EM}}
\newcommand{\Fraisse}{Fra\"{i}ss\'{e}}
\newcommand{\Lovasz}{Lov\'{a}sz}
\newcommand{\kcirc}{\circ_{\Kl}}

\begin{document}

\title[The Pebble-Relation Comonad in Finite Model Theory]{The Pebble-Relation Comonad in Finite Model Theory}

\author[Y.~Montacute]{Yo\`av Montacute\lmcsorcid{0000-0001-9814-7323}}[a]
\author[N.~Shah]{Nihil Shah\lmcsorcid{0000-0003-2844-0828}}[b]

\address{University of Cambridge, United Kingdom}	

\email{yoav.montacute@cl.cam.ac.uk}  

\address{University of Oxford, United Kingdom}	
\email{nihil.shah@cs.ox.ac.uk}  

\begin{abstract}
The pebbling comonad, introduced by Abramsky, Dawar and Wang, provides a categorical interpretation for the $k$-pebble games from finite model theory. The coKleisli category of the pebbling comonad specifies equivalences under different fragments and extensions of infinitary $k$-variable logic. Moreover, the coalgebras over this pebbling comonad characterise treewidth and correspond to tree decompositions. 
In this paper we introduce the pebble-relation comonad, which characterises pathwidth and whose coalgebras correspond to path decompositions. We further show that the existence of a coKleisli morphism in this comonad is equivalent to truth preservation in the restricted conjunction fragment of $k$-variable infinitary logic. We do this using Dalmau's pebble-relation game and an equivalent all-in-one pebble game. We then provide a similar treatment to the corresponding coKleisli isomorphisms via a bijective version of the all-in-one pebble game. Finally, we show as a consequence a new {\Lovasz}-type theorem relating pathwidth to the restricted conjunction fragment of $k$-variable infinitary logic with counting quantifiers.
\end{abstract}

\maketitle


\section{Introduction}
Model theory is a field in which mathematical structures are not seen as they really are, i.e.\ up to isomorphism, but through the fuzzy reflection imposed by their definability in some logic $\Logic$. 
Namely, given two structures over the same signature $\As$ and $\Bs$, model theory is concerned with equivalence under the relation
\[\As \equivL \Bs = \forall \phi \in \Logic, \; \As \models \phi \IFF \Bs \models \phi.\]
Historically, a central theme in model theory has been to find syntax-free characterisations of these equivalences. 
This is exemplified by the Keisler-Shelah theorem \cite{shelah1971} for first-order logic. 
These equivalences are also characterised by model-comparison, or Spoiler-Duplicator games. 
Spoiler-Duplicator games are graded by a (typically finite) ordinal that corresponds to a grading of some syntactic resource. 
For example, the $k$-pebble game introduced by Immerman \cite{immerman1982} characterises equivalence in infinitary logic graded by the number of variables $k$.

Abramsky, Dawar and Wang \cite{abramsky2017} provided this game, and two similar variants, with a categorical interpretation in terms of morphisms which involve the pebbling comonad $\Pk$. 
Since then similar game comonads were discovered for the Ehrenfeucht-{\Fraisse} games \cite{abramskyResources2018}, modal-bisimulation games \cite{abramskyResources2018}, games for guarded \mbox{logics \cite{abramsky2020}} and games for finite-variable logics with generalised \mbox{quantifiers \cite{oconghaile2020}}.
In all of these cases the coalgebras over the game comonad correspond to decompositions of structures such as tree decompositions of width $< k$ and forest covers of height $\leq k$, in the cases of the pebbling comonad and the Ehrenfeucht-{\Fraisse} comonad respectively. 
As an immediate corollary of these correspondences, we obtain alternative and novel definitions for associated graph parameters such as treewidth in the case of the pebbling comonad and tree-depth in the case of the Ehrenfeucht-{\Fraisse} comonad. 

 Abramsky et al.\ \cite{abramsky2017} proved that the coKleisli morphisms associated with $\Pk$ correspond to Duplicator's winning strategies in the one-sided $k$-pebble game. 
 This one-sided game, introduced by Kolaitis and Vardi \cite{kolaitis1990}, was used to study expressivity in \textsf{Datalog}, preservation of existential positive sentences of $k$-variable logic $\pLogicW$ and \textbf{P}-tractable constraint satisfaction problems. 
 \mbox{Dalmau \cite{dalmau2005}} developed an analogous one-sided pebble-relation game that he used to study expressivity in linear \textsf{Datalog}, preservation of a restricted conjunction fragment of existential positive $k$-variable logic $\pRLogicK$ and \textbf{NL}-tractable constraint satisfaction problems. 
Further, Abramsky et al.\ \cite{abramsky2017} demonstrated that, for the pebbling comonad, the coKleisli isomorphism associated with $\Pk$ corresponds to Hella's bijective variant of the $k$-pebble game \cite{hella1996}. However, the bijective version of the pebble-relation game has not been explored.

In this paper we widen the domain of these game comonads by introducing the pebble-relation comonad $\Rk$, where the coalgebras correspond to path decompositions of width $< k$. 
This yields a new definition for pathwidth.
Moreover, just as coKleisli morphisms of $\Pk$ correspond to Duplicator's winning strategies in the one-sided $k$-pebble game, we show that coKleisli morphisms of $\Rk$ correspond to Duplicator's winning strategies in Dalmau's $k$-pebble relation game \cite{dalmau2005}.
We do this by introducing an equivalent game which, as will be detailed later, is much more suitable for our purpose. 
This game is called the \emph{all-in-one $k$-pebble game} and the existence of a winning strategy for Duplicator in this game is equivalent to truth preservation in the restricted conjunction fragment of $k$-variable infinitary logic.
Inspired by the isomorphism result for $\Pk$, we define a bijective variant of this game.
Consequently, the coKleisli isomorphism allows us to obtain a new characterisation of equivalence in the, heretofore unexplored, restricted conjunction fragment of $k$-variable infinitary logic with counting quantifiers. 
It is worth mentioning that equivalence in this fragment may yield a new isomorphism approximation method \cite{weisfeiler1968}, similarly to how equivalences in $k+1$-variable counting logic can be algorithmically interpreted as graphs which are indistinguishable by the polynomial time $k$-Weisfeiler-Lehman isomorphism approximation \mbox{method \cite{grohe2017}}.

As one immediate consequence of this comonadic interpretation, we will demonstrate that equivalence in the restricted conjunction fragment of $k$-variable logic with counting quantifiers is equivalent to counting homomorphisms from structures of pathwidth $< k$.
This applies and expands the theoretical picture established in a recent paper by Dawar, Jakl and Reggio \cite{dawar2021} regarding a categorical {\Lovasz}-type theorem. 
It is also intertwined with the growing interest in the relationship between homomorphism counting and pathwidth as discussed for example in Dell, Grohe and Ratten \cite{holger2018}.

\paragraph{Outline} Section \ref{sec:background} introduces the necessary preliminaries and notation we use throughout the paper. 
It includes the necessary background on Spoiler-Duplicator game comonads and the associated results that accompany them.
Section \ref{sec:ourComonad} discusses the pebble-relation comonad $\Rk$ and demonstrates the relationship between $\Rk$ and $\Pk$. 
Section \ref{sec:coalgebras} proves that coalgebras over $\Rk$ correspond to path decompositions of width $< k$, which provides a coalgebraic characterisation of pathwidth. 
Section \ref{sec:equivalences} introduces Dalmau's one-sided pebble-relation game and the equivalent all-in-one pebble game in which Duplicator's winning strategies are captured through morphisms involving $\Rk$.
The same treatment is then given to the bijective version of the all-in-one pebble game in which Duplicator's winning strategies are captured by isomorphisms involving $\Rk$. 
The section concludes with issuing a {\Lovasz}-type theorem for pathwidth. 
Section \ref{sec:conclusion} concludes the paper with a summary of the results and a discussion on further research directions involving the pebble-relation comonad. 
\section{Preliminaries}\label{sec:background}
In this section, we will establish some notational preliminaries and provide a short introduction to the relevant concepts in category theory and finite model theory used throughout the paper.
\subsection{Set notation}
Given a partially ordered set $(X,\leq)$ and $x \in X$, the \emph{down-set} and \emph{up-set} of $x$ are ${\darr x} = \{y \in X \mid y \leq x\}$ and ${\uarr x} = \{y \in X \mid y \geq x\}$ respectively.  
A partially ordered set $(X,\leq)$ is a \textit{linear order}, or a \textit{directed path}, if for every pair of elements $x,y \in X$, $x \leq y$ or $y \leq x$, i.e.\ every two elements are comparable.
If $(T,\leq)$ is a partially ordered set such that for every $x \in T$, ${\darr x}$ is linearly ordered by $\leq$, then $\leq$ \textit{ forest orders $T$} and $(T,\leq)$ is a \textit{forest}. 
The \emph{height of an element $x \in T$} of a forest $(T,\leq)$ is the cardinality of ${\downarrow x} \backslash \{x\}$.
The \emph{height of a forest $(T,\leq)$} is the maximal height of an element $x \in T$.
If $(T,\leq)$ is a forest and there exists a least element $\bot \in T$ such that, for all $x \in T$, $\bot \leq x$, then \textit{$\leq$ tree orders $T$} and $(T,\leq)$ is a \textit{tree}. 
For the purpose of forest orders, we will use interval notation, i.e.\ $(x,x'] = \{y \mid x < y \leq x'\}$ and $[x,x'] = \{y \mid x \leq y \leq x'\}$.

For a positive integer $n$, we denote by $\setn$ the set $\{1,\dots,n\}$. 
As a convention, we consider $\setn$ as having the usual order on segments of natural numbers $\leq$.
Given a set $A$, we denote the set of finite sequences of elements $A$ as $A^{*}$ and non-empty finite sequences as $A^{+}$.
The set of sequences of length $\leq k$ is denoted by $A^{\leq k}$. 
We denote the sequence of elements $a_1,\dots,a_n \in A$ as $[a_1,\dots,a_n]$ and the empty sequence as $\emptyseq$. 
We write $|s| = n$ for the length of a sequence $s = [a_1,\dots,a_n]$. 
Given two sequences $s,t \in A^{*}$, we denote the concatenation of $s$ followed by $t$ as $st$. 
If $s$ is such that there exists a (possibly empty) sequence $s'$ where $ss' = t$, then we write $s \sqsubseteq t$. 
Observe that $\sqsubseteq$ defines a relation on sequences and tree orders $A^{*}$ and forest orders $A^{+}$. 
Hence, by the interval notation defined for forest orders, if $s \sqsubseteq t$, then $(s,t]$ denotes the suffix of $s$ in $t$. 
For $s = [a_1,\dots,a_n]$ and $i,j \in \setn$, we define a range notation where $s(i,j] = [a_{i+1},\dots,a_j]$ if $i < j$ or $\emptyseq$ if $i \geq j$, and $s[i,j] = [a_i,\dots,a_j]$ if $i \leq j$ or $\emptyseq$ if $i > j$.

\subsection{Category theory}
We assume familiarity with the standard category-theoretic notions of category, functor and natural transformation.
Given a \mbox{category $\mathfrak{C}$}, we denote the class of its objects by $\mathfrak{C}_0$ and the class of its morphisms by $\mathfrak{C}_1$.
We define the notion of a relative comonad which weakens the endofunctor requirement of the standard comonad. The relative comonad is the dual of the \emph{relative monad} introduced in \cite{altenkirch2014}.  
Given a functor $J:\mathfrak{J} \rightarrow \mathfrak{C}$, a \textit{relative comonad on $J$} is a triple $(\mathbb{T},\varepsilon,(\cdot)^{*})$, where $\T:\mathfrak{J}_0 \rightarrow \mathfrak{C}_0$ is an object mapping, $\varepsilon_{X}:\T X \rightarrow J X \in \mathfrak{C}_1$ is defined for every object $X \in \mathfrak{J}_0$ and $f^{*}:\T X \rightarrow \T Y \in \mathfrak{C}_1$ is a coextension map, for each $f:\T X \rightarrow J Y \in \mathfrak{C}_1$, satisfying the following equations: 
\[ \varepsilon_{X}^* = \id_{\T X}; \qquad \varepsilon \circ f^* = f; \qquad (g \circ f^*)^* = g^* \circ f^* . \]
These equations allow us to extend the object mapping $\T$ to a functor, where $\T f = (Jf \circ \varepsilon_{X})^{*}$ for $f:X \rightarrow Y$. 
For every relative comonad $(\T,\varepsilon,(\cdot)^{*})$ over $J:\mathfrak{J} \rightarrow \mathfrak{C}$, we can define an associated \textit{coKleisli category}, denoted by $\Kl(\T)$, where
\begin{itemize}
  \item $\Kl(\T)_{0}$ is the same as the class of objects $\mathfrak{J}_0$;
  \item $\Kl(\T)_{1}$ are morphisms of type $f:\T X \rightarrow J Y \in \mathfrak{C}_1$;
  \item The composition $g \kcirc f:\T X \rightarrow J Z$ of  two morphisms $f:\T X \rightarrow J Y$ and $g: \T Y \rightarrow J Z$ is given by
    \[\T X \xrightarrow{f^{*}} \T Y \xrightarrow{g} J Z;\]
  \item The identity morphisms are given by the counit  \[\varepsilon_{X}:\T X \rightarrow J X.\]
\end{itemize}

The ordinary notion of a \textit{comonad in coKleisli form} \cite{manes1976} and the corresponding Kleisli category can be recovered when $\mathfrak{J} = \mathfrak{C}$ and $J = \textsf{id}_{\mathfrak{C}}$. 
Observe that given a comonad $(\T:\mathfrak{C} \rightarrow \mathfrak{C},\varepsilon,(\cdot)^{*})$ and a functor $J:\mathfrak{J} \rightarrow \mathfrak{C}$, the functor $\T^{J} = \T \circ J$ can be made into a relative comonad $(\T^{J},\varepsilon',(\cdot)')$ on $J$, where $\varepsilon'_{X} = \varepsilon_{J X}$ and the coextension mapping $()^{'}$ is defined for $f:\T^{J} X \rightarrow J Y$ to be the same as $f^{*}$, i.e.\ $f^{'} = f^{*}: \T^{J} \rightarrow \T^{J} Y$.

Given an ordinary comonad in coKleisli form, we can define a comultiplication morphism $\delta_{X}:\T X \rightarrow \T \T X$, where $\delta_{X} = (\textsf{id}_{\T X})^{*}$, which satisfies the following equations:
\[ \T \delta_{X} \circ \delta_{X} = \delta_{\T X} \circ \delta_{X}; \qquad \T \varepsilon_{X} \circ \delta_{X} = \varepsilon_{\T X} \circ \delta_{X} = \id_{\T X}. \]
The triple $(\T:\mathfrak{C} \rightarrow \mathfrak{C},\varepsilon,\delta)$, where $\T$ is a functor, is a \textit{comonad in standard form}. 
Its coKleisli form can be recovered by defining the coextension mapping $(\cdot)^{*}$ as $f^{*} = \T f \circ \delta$.

Given a comonad $(\T:\mathfrak{C} \rightarrow \mathfrak{C},\varepsilon,\delta)$, a \textit{coalgebra} over $\T$ is defined as a pair $(A,\alpha:A \rightarrow \T A)$, where $A \in \mathfrak{C}_0$ and $\alpha \in \mathfrak{C}_1$, satisfying the following equations: 
\[ \delta_{A} \circ \alpha = \T \alpha \circ \alpha; \qquad \varepsilon_{A} \circ \alpha = \id_{A}.  \]
The \textit{category of coalgebras}, or \textit{Eilenberg-Moore category}, associated with a comonad $\T:\mathfrak{C} \rightarrow \mathfrak{C}$, denoted by $\EM(\T)$, is defined as
\begin{itemize}
    \item $\EM(\T)_0$ consists of coalgebras $(A,\alpha:A \rightarrow \T A)$;
    \item A morphism of type $h:(A,\alpha) \rightarrow (B,\beta) \in \EM(\T)_1$ is a morphism $h:A \rightarrow B \in \mathfrak{C}_1$ such that $\T h \circ \alpha = \beta \circ h$;
    \item Identity and composition are defined as in $\mathfrak{C}$.
\end{itemize}
A \textit{resolution} of $\T$ is an adjunction $L:\mathfrak{C} \rightarrow \mathfrak{D} \dashv R:\mathfrak{D} \rightarrow \mathfrak{C}$ that gives rise to $\T$ as a comonad. 
A resolution $L \dashv R$ is a \textit{comonadic adjunction} if the comparison functor $K:\mathfrak{D} \rightarrow \EM(\T)$ is an equivalence.

\subsection{Finite model theory}
We fix a vocabulary $\sigma$ of relational symbols $R$, each with a positive arity $\rho(R)$.
If $R$ has arity $\rho(R)$, then $R$ is called an \textit{$\rho(R)$-ary relation}.
A \textit{$\sigma$-structure} $\As$ is specified by a universe of elements $A$ and interpretations $R^{\As} \subseteq A^{\rho(R)}$ for each relation $R \in \sigma$. 
We use calligraphic letters ($\As,\Bs,\Cs$, etc.) to denote $\sigma$-structures and we use roman letters ($A,B,C$, etc.) to denote their underlying universes of elements. 

Let $\As$ and $\Bs$ be $\sigma$-structures. 
If $B \subseteq A$ and $R^{\Bs} \subseteq R^{\As}$ for every relation $R \in \sigma$, then $\Bs$ is a \textit{$\sigma$-substructure} of $\As$.
If $B \subseteq A$, then we can form the \textit{$B$ induced $\sigma$-substructure} $\Bs = \As|_{B}$ with universe $B$ and interpretations $R^{\Bs} = R^{\As} \cap B^{\rho(R)}$ for each relation $R \in \sigma$. 
The graph $\Gf(\As) = (A,\arc)$ is the Gaifman graph of $\As$, where $a \arc a'$ iff $a = a'$ or $a,a'$ appear in some tuple of $R^{\As}$ for some $R \in \sigma$.

A \textit{$\sigma$-morphism from $\As$ to $\Bs$}, or \textit{homomorphism}, $h:\As \rightarrow \Bs$ is a set function from $A$ to $B$ such that $R^{\As}(a_1,\dots,a_{\rho(R)})$ implies $R^{\Bs}(h(a_1),\dots,h(a_{\rho(R)}))$ for every relation $R \in \sigma$. 
We will denote the category of $\sigma$-structures with $\sigma$-morphisms by $\Rsig$. 
The full subcategory of $\sigma$-structures with finite universes is denoted by $\Rsigf$. 
If there exists a homomorphism $h:\As \rightarrow \Bs$, we write $\As \rightarrow \Bs$. 
In the category $\Rsig$, an isomorphism $f:\As \rightarrow \Bs$ is a bijective homomorphism that reflects relations, i.e.\ for each relation $R\in\sigma$,
\[R^{\As}(a_1,\dots,a_m) \Leftrightarrow R^{\Bs}(f(a_1),\dots,f(a_{\rho(R)})).\]
\subsection{Logical fragments}
We are mainly concerned with fragments of infinitary logic.
The infinitary logic $\LogicW$ has the standard syntax and semantics of first-order logic, but where disjunctions and conjunctions are allowed to be taken over arbitrary sets of formulas. 
We denote formulas with free variables among $\vec{x} = (x_1,\dots,x_n)$ by $\phi(\vec{x})$. 
If $\As$ is a $\sigma$-structure, $\vec{a} \in A^{n}$, $\phi(\vec{x}) \in \LogicW$ and $\As,\vec{a}$ satisfies $\phi(\vec{x})$, then we write $\As,\vec{a} \models \phi(\vec{x})$.

For vectors of variables and elements $\vec{x} = (x_1,\dots,x_n)$, we write $[\vec{x}] = \{x_1,\dots,x_n\}$ for the underlying \textit{support} of the vector.
The infinitary logic can be graded into $k$-variable fragments, denoted by $\LogicK$, where formulas contain at most $k$-many variables.  

For every logic $\Logic$ considered throughout the paper, we will also be interested in two variants. 
The first variant is the existential positive fragment $\pLogic$, where we only consider formulas constructed using existential quantifiers, disjunctions, conjunctions and atomic formulas. 
The second variant involves a restricted conjunction. 
A \textit{restricted conjunction} is a conjunction of the form $\bigwedge \Psi$, where $\Psi$ is a set of formulas satisfying the following condition:
\begin{enumerate}[label=(R)]
    \item \label{linearCond} At most one formula $\psi\in\Psi$ with quantifiers is not a sentence. 
\end{enumerate}
The main goal of this paper is to study the \textit{restricted conjunction logics} $\pRLogicK$ and $\cRLogicK$ (Section \ref{newlogic}).

Given two $\sigma$-structures $\As$ and $\Bs$, if for all sentences $\phi \in \Logic(\sigma)$, $\As \models \phi $ implies $\Bs \models \phi$, then we write $\As \presL \Bs$. 
If $\As \presL \Bs$ and $\Bs \presL \As$, then we write $\As \equivL \Bs$. 
For logics $\Logic$ closed under negation, we have that $\As \presL \Bs$ implies $\As \equivL \Bs$.

\subsection{Spoiler-Duplicator games}
The relations $\presL$ and $\equivL$ for specific choices of $\Logic$ are characterised, in a syntax-free fashion, by Spoiler-Duplicator games (also called model-comparison games or Ehrenfeucht-{\Fraisse} style games). 
In each game, we consider two $\sigma$-structures, $\As$ and $\Bs$, and two players: Spoiler, who tries to show that two structures are different under $\Logic$; and Duplicator, who tries to show the two structures are the same under $\Logic$ (i.e.\ $\As \equivL \Bs$). 
The standard Ehrenfeucht-{\Fraisse} game appears in many introductory texts on model theory (see e.g.\ \cite{libkin}). 
The model-comparison games we are interested in are modified versions of the one-sided $k$-pebble game \cite{immerman1982} and the bijective $k$-pebble game \cite{hella1996}. 
Each game is played in a number of rounds.

 For the one-sided $k$-pebble game $\exists^{+}\Peb_k(\As,\Bs)$, which characterises $\Rrightarrow^{\LogicK}$, both Spoiler and Duplicator have a set of $k$ pebbles. At each $n$-th round such that $n \in \omega$:
    \begin{itemize}
      \item Spoiler places a pebble $p_n \in \setk$ on an element $a_n \in A$. If the pebble $p_n \in \setk$ is already placed on an element of $A$, Spoiler moves the pebble from that element to the chosen element;
      \item Duplicator places the pebble $p_n \in \setk$ on an element $b_n \in B$. 
    \end{itemize}
    At the end of the $n$-th round, we have a pair of sequences $s = [(p_1,a_1),\dots,(p_n,a_n)]$ and $ t =[(p_1,b_1),\dots,(p_n,b_n)]$.
    For every $p$ in the set $\pebbles(s) = \{p_1,\dots,p_n\}$ of pebbles appearing in $s$, let $a^{p} = \last_{p}(s)$ and $b^{p} = \last_{p}(t)$ be the last elements pebbled with $p$ in $s$ and $t$ respectively. 
    If the relation $\gamma_{n} = \{(a^{p},b^{p}) \mid p \in \pebbles(s) \}$ is a partial homomorphism from $\As$ to $\Bs$, then Duplicator wins the $n$-th round of the $k$-pebble game.\footnote{When Spoiler (resp. Duplicator) makes a choice that results in her winning a round, we often say that this choice is a \textit{winning move} for her.}
    Duplicator has a winning strategy in $\exists^{+}\Peb_k(\As,\Bs)$ if for every round $n \in \omega$, and for every move by Spoiler in the $n$-th round, Duplicator has a winning move.
    
  For the bijective $k$-pebble game $\# \Peb_k(\As,\Bs)$, Spoiler wins automatically if there is no bijection between $A$ and $B$. At each $n$-th round such that $n \in \omega$:
    \begin{itemize}
        \item Spoiler chooses a pebble $p \in \setk$;
        \item Duplicator responds with a bijection $f_n:A \rightarrow B$ consistent with the previously placed pebbles, i.e.\ for every $q \not= p$ such that $(a^{q},b^{q}) \in \gamma_{n-1}$, $f_n(a^{q}) = b^{q}$;
        \item Spoiler places the pebble $p$ on an element $a \in A$. Duplicator places the pebble $p$ on the element $f(a) \in B$.
    \end{itemize}
    Duplicator wins the $n$-th round of $\# \Peb_k(\As,\Bs)$ if the relation $\gamma_n = \{(a^{p},f_n(a^{p})) \mid p \in \setk\}$ is a partial isomorphism. As before, Duplicator has a winning strategy if she can keep playing forever. 

Capturing these games as constructions on the category of relational structures has been the underlying theme of the research program motivating this paper.

\subsection{Spoiler-Duplicator game comonads}
\label{sec:comonads}
The larger research program motivating this paper is the discovery of Spoiler-Duplicator game comonads $\CC_k$ associated with logics $\Logic$ graded by a resource $k$.
Though these Spoiler-Duplicator game comonads have no general definition that applies to all cases, different indexed families of comonads $\CC_k$ over $\Rsig$ exhibit the same pattern of results: 
 the morphism power theorem, the isomorphism power theorem and the coalgebra characterisation theorem  \cite{abramsky2017,abramskyResources2018,oconghaile2020,abramsky2020}. 
In this section, we will review the general schema for each of these results.  

\paragraph{Morphism and Isomorphism Power Theorems}
Given a logic $\Logic_k$, graded by some syntactic resource $k$ (e.g.\ number of variables, quantifier rank and modal depth) and corresponding model-comparison game $\GG_k$, the Spoiler-Duplicator comonad associated with $\GG_k$ is an indexed family of comonads $\CC_k$ over $\Rsig$ \cite{abramskyResources2018}. 
We can then leverage the coKleisli category $\Kl(\CC_k)$ associated with $\CC_k$ to capture the relation $\Rrightarrow^{\pLogic_k}$ of the sentences in the existential positive fragment of $\Logic_k$ limited by the resource $k$. 
\begin{thm}[Morphism Power Theorem]
For all {$\sigma$-structures} $\As$ and $\Bs$, the following are equivalent:
\begin{enumerate}
  \item Duplicator has a winning strategy in $\exists^{+} \GG_k(\As,\Bs)$.
  \item $\As \Rrightarrow^{\pLogic_k} \Bs$.
  \item There exists a coKleisli morphism $f:\CC_k \As \rightarrow \Bs$.
\end{enumerate}
\end{thm}
For example, in the case of the pebbling comonad $\Pk$, a coKleisli morphism $\Pk \As \rightarrow \Bs$ corresponds to a winning strategy for Duplicator in $\exists^{+}\Peb_k(\As,\Bs)$ and characterises the relation $\Rrightarrow^{\pLogicK}$ \cite{abramsky2017}.
The indexing of $\CC_k$ models resources in the corresponding game. 
Namely, for $k \leq l$, there is a comonad inclusion $\CC_k \hookrightarrow \mathbb{C}_{l}$. 
Interpreting the inclusion in terms of the game corresponds to the fact that Spoiler playing with $k$ resources is a special case of Spoiler playing with $l \geq k$ resources. 
This means that the smaller the $k$ is, the easier it is to find morphisms of type $\CC_k \As \rightarrow \Bs$. 

Ostensibly, the asymmetry of a coKleisli morphism means that only the ``forth'' aspect of the game $\GG_k$ (the game $\exists^{+} \GG_k$) can be captured by this comonadic approach. 
A natural candidate for capturing the symmetric game would be to consider the symmetric relation of coKleisli isomorphism, i.e.\ there exist morphisms $f:\CC_k \As \rightarrow \Bs$ and $g:\CC_k \Bs \rightarrow \As$ such that $g \kcirc f = \varepsilon_{\As}$ and $f \kcirc g = \varepsilon_{\Bs}$.
However, the isomorphisms in $\Kl(\CC_k)$ characterise the equivalence relation $\equiv^{\cLogic_k}$ for the logic $\Logic_k$ extended with counting quantifiers. This results in a logic stronger than $\Logic_k$. 

\begin{thm}[Isomorphism Power Theorem]
For all finite $\sigma$-structures $\As$ and $\Bs$, the following are equivalent:
\begin{enumerate}
  \item Duplicator has a winning strategy in $\# \GG_k(\As,\Bs)$.
  \item $\As \equiv^{\cLogic_k} \Bs$.
  \item There exists a coKleisli isomorphism $f:\CC_k \As \rightarrow \Bs$.
\end{enumerate}
\end{thm}
For example, in the case of the pebbling comonad $\Pk$, a coKleisli isomorphism $\Pk \As \rightarrow \Bs$ corresponds to a winning strategy for Duplicator in $\#\Peb_k(\As,\Bs)$ and characterises the relation $\equiv^{\cLogicK}$ \cite{abramsky2017}.
\paragraph{Coalgebras and Adjunctions}
    A natural inquiry regarding the comonads $\CC_k$ is to investigate the category of coalgebras $\EM(\CC_k)$. 
    It turns out that for all the cases of $\CC_k$ constructed from some game $\GG_k$, the coalgebras $\As \rightarrow \CC_k \As$ correspond to forest covers of structures $\As$.
    
    Define a \textit{forest cover for $\As$} to be a forest $(S,\leq)$, where $A \subseteq S$, such that if $a \frown a' \in \As$, i.e.\ $a$ and $a'$ are related in some tuple of $\As$, then $a \leq a'$ or $a' \leq a$. In this paper, we assume that all forest covers are \textit{tight} forest covers, i.e.\ where $S = A$.
\begin{thm}[coalgebra characterisation]
The following statements are equivalent:
\begin{enumerate}
  \item $\As$ has a forest cover of parameter $\leq k$.
  \item There exists a coalgebra $\alpha:\As \rightarrow \CC_k \As$.
\end{enumerate}
\end{thm}

For every game comonad $\CC_k$, we define the \textit{$\mathbb{C}$-coalgebra number of $\As$} to be the least $k$ such that there exists a coalgebra $\As \rightarrow \CC_k \As$. We denote it by $\kappa^{\mathbb{C}}(\As)$.  
This definition together with the coalgebra characterisation theorems allow us to obtain new alternative definitions for various combinatorial invariants of relational structures. 

For example, in the case of the Ehrenfeucht-{\Fraisse} comonad $\Ek$, a coalgebra over $\As$ corresponds to a forest cover of $\As$ with height $\leq k$. 
Consequently, $\kappa^{\mathbb{E}}(\As)$ is equal to the tree-depth of $\As$ \cite{abramskyResources2018}. 
Moreover, in the case of the pebbling comonad, a coalgebra over $\As$ corresponds to a forest cover over $\As$ with an additional pebbling function $p:A \rightarrow \setk$ that encodes the data of a tree decomposition for $\As$ of width $<k$. Consequently, $\kappa^{\mathbb{P}}(\As)$ corresponds to the treewidth of $\As$ \cite{abramsky2017}. 

The bijective correspondence between forest covers of a certain type and coalgebras can be extended to isomorphisms of the respective categories.
In the language of adjunctions, for every $\CC_k$, we can form the category of forest-ordered $\sigma$-structures $\TsigCk$ consisting of
\begin{itemize}
    \item $\TsigCk_0$ are pairs $(\As,\leq)$, where $\As \in \Rsig_0$ and $\leq$ forest-orders the universe $A$ of $\As$ such that the following condition holds: 
    \begin{center}
    (E) if $a \arc a' \in \As$, then $a \leq a'$ or $a' \leq a$.
    \end{center}
    \item $\TsigCk_1$ are $\sigma$-morphisms $f:\As \rightarrow \Bs \in \Rsig_1$ that preserve roots and the covering relation of $\leq$. That is, $f$ preserves immediate successors in $\leq$. 
    \item The identity and composition are inherited from $\Rsig$.
\end{itemize}
Evidently, there exists a forgetful functor $U_k:\TsigCk \rightarrow \Rsig$.
We may view $\CC_k$ as the comonad arising from constructing a functor $F_k:\Rsig \rightarrow \TsigCk$ that is right adjoint to $U_k$ and where the adjunction $U_k \dashv F_k$ is comonadic. 
That is, $\TsigCk$ and $\EM(\CC_k)$ are equivalent categories.

For example, the category of coalgebras for the Ehrenfeucht-{\Fraisse} comonad is equivalent to the category of forest covers of height $\leq k$ \cite{abramskyResources2021}. 
Similarly, the category of coalgebras $\EM(\Pk)$ for the pebbling comonad is equivalent to the category of $k$-pebble forest covers \cite{abramskyResources2021}.

For each of the result schemata (morphism power theorem, isomorphism power theorem and coalgebra characterisation theorem), we mentioned the corresponding result for the pebbling comonad. 
However, we would like to stress that analogous results hold for all of the comonads previously discussed in the literature \cite{abramsky2017,abramskyResources2018,abramsky2020,oconghaile2020}. Obtaining these results from a general definition is an active topic of research.

\section{Pebble-Relation Comonad}
\label{sec:ourComonad}
The pebble-relation comonad, which captures pathwidth, is closely related to the pebbling comonad $(\Pk,\varepsilon,()^{*})$ that is used to define treewidth \cite{abramsky2017,abramskyResources2018}.
 Given a $\sigma$-structure $\As$, we define the universe of $\Pk \As$ as $(\setk \times A)^{+}$. 
The counit morphism $\varepsilon_{\As}:\Pk\As \rightarrow \As$ is defined as the second component of the last element of the sequence, i.e.\ $\varepsilon_{\As}[(p_1,a_1),\dots,(p_n,a_n)] = a_n$. 
The coextension of a morphism $f:\Pk \As \rightarrow \Bs$ is defined as \[f^{*}[(p_1,a_1),\dots,(p_n,a_n)] = [(p_1,b_1),\dots,(p_n,b_n)],\] where $b_i = f[(p_1,a_1),\dots,(p_i,a_i)]$, for all $i \in \setn$. 

To define the $\sigma$-structure on $\Pk\As$, we will need to introduce some new notation. 
The mapping $\pi_{A}:\Pk A \rightarrow \setk$ is defined as the first component of the last element of the sequence, i.e.\ $\pi_{A}[(p_1,a_1),\dots,(p_n,a_n)] = p_n$. 
Suppose $R \in \sigma$ is an $m$-ary relation. Then $R^{\Pk \As}(s_1,\dots,s_m)$ iff
  \begin{enumerate}
        \item $\forall i,j \in \setm, s_i \sqsubseteq s_j$ or $s_j \sqsubseteq s_i$; \hfill (\emph{pairwise comparability})
        \item If $s_i\sqsubseteq s_j$, then $\pi_{A}(s_i)$ is not in $(s_i,s_j]$,\\ and similarly for $s_j\sqsubseteq s_i$;\hfill (\emph{active pebble})
        \item $R^{\As}(\varepsilon_{\As}(s_1),\dots,\varepsilon_{\As}(s_m))$. \hfill (\emph{compatibility})
\end{enumerate}

The elements of $\Pk\As$ can be seen as the set of Spoiler plays in the $k$-pebble game on the structure $\As$. 
The additional active pebble condition models how Spoiler's choice of $k$ pebble placements amounts to moving a $k$-sized window of variables that are assigned to elements in the structure $\As$. 
When working with $\Pk\As$, as before, we will use $\pebbles(s)$ to denote the set of pebbles appearing in $s$, $\last_{p}(s)$ to denote the last element $a \in \As$ in the sequence $s$ pebbled by $p \in \pebbles(s)$. 
The set of \emph{active elements of $s$} is defined as $\Active(s) = \{\last_{p}(s) \mid p \in \pebbles(s) \}$. 
Given two sequences $s \in \Pk A$ and $t \in \Pk B$, we define the relation of pebbled elements $\gamma_{s,t} = \{(\last_{p}(s),\last_{p}(t))\} \subseteq A \times B$.

We now introduce our main construction, which is the family of comonads $(\Rk,\varepsilon,()^{*})$, for every $k \in \omega$ over $\Rsig$. 
Given a $\sigma$-structure $\As$,  we define the universe of $\Rk \As$ as
\[\Rk A = \{([(p_1,a_1),\dots,(p_n,a_n)],i)\},\] where  $(p_j,a_j) \in \setk \times A$ and $i \in \setn$.
Intuitively, $\Rk A$ is the set of Spoiler plays in the $k$-pebble game paired with an index denoting a move of the play.
The counit morphism $\varepsilon_{\As}:\Rk \As \rightarrow \As$ is defined as $\varepsilon_{\As}([(p_1,a_1),\dots,(p_n,a_n)],i) = a_i$. 
The coextension of a morphism $f:\Rk \As \rightarrow \Bs$ is defined as \[f^{*}([(p_1,a_1),\dots,(p_n,a_n)],i) = ([(p_1,b_1),\dots,(p_n,b_n)],i),\] where $b_j = f([(p_1,a_1),\dots,(p_n,a_n)],j)$ for all $j \in \setn$.

We can interpret the relations on $\Rk\As$ in a similar manner to the interpretations given for $\Pk\As$:
Suppose $R \in \sigma$ is an $m$-ary relation. Then  $R^{\Rk \As}((s_1,i_1),\dots,(s_m,i_m))$ iff
\begin{enumerate}
    \item $\forall j \in \setm, s_j = s$; \hfill (\emph{equality})
     \item $\pi_{A}(s,i_j)$ does not appear in $s(i_j,i]$; \hfill (\emph{active pebble})
    \item $R^{\As}(\varepsilon_{\As}(s,i_1),\dots,\varepsilon_{\As}(s,i_m))$; \hfill (\emph{compatibility})
\end{enumerate}
where $i = \max\{i_1,\dots,i_m\}$ and $\pi_{A}:A \rightarrow \setk$ is defined as \[\pi_{A}([(p_1,a_1),\dots,(p_n,a_n)],i) = p_i.\] 
The active pebble and compatibility conditions play a similar role as in the definition of $\Pk\As$. 
As we will see in Section \ref{sec:equivalences}, the equality condition ensures that Duplicator must respond to a (full) Spoiler play in one round. 

\begin{prop}
  $(\Rk,\varepsilon,()^{*})$ is a comonad in coKleisli form.
\end{prop}
\begin{proof}
  It is easy to see that $\Rk$ is a lifting of the pointed-list (see e.g.\ \cite{orchardThesis}) comonad over \textbf{Set}. 
  In order to show that $\Rk$ is a comonad over $\Rsig$, we need to show that $\varepsilon_{\As}$ is a $\sigma$-morphism and that for every $f \in \Rsig_1$, $f^{*} \in \Rsig_1$.
  The fact that $\varepsilon_{\As}$ is a $\sigma$-morphism follows from the compatibility condition in the definition of $R^{\Rk\As}$. 
  To show that $f^{*}$ is a $\sigma$-morphism, suppose that $R^{\Rk\As}((s,i_1),\dots,(s,i_m))$ and that $s = [(p_1,a_1),\dots,(p_n,a_n)]$. 
  Consider $t = [(p_1,b_1),\dots,(p_n,b_n)]$, where $f(s,i) = b_i$. It follows that $f^{*}(s,i) = (t,i)$.
  By construction, $\pi_{\As}(s,i_j) = \pi_{\Bs}(t,i_j)$, so $((t,i_1),\dots,(t,i_m))$ satisfies the active pebble condition. 
  Since $f$ is a $\sigma$-morphism, $R^{\Rk\As}((s,i_1),\dots,(s,i_m))$ and $\varepsilon_{\Bs}(t,i) = b_i$, we have that $R^{\Bs}(b_{i_1},\dots,b_{i_m})$. 
  Therefore, the compatibility condition holds, $R^{\Rk\Bs}((t,i_1),\dots,(t,i_m))$ and $f^{*}$ is a $\sigma$-morphism.
\end{proof}

We also define for every $n \in \omega$, $\Rkn\As \subseteq \Rk\As$ to be the substructure induced by pairs $(s,i)$, where $s$ is of length $\leq n$. 
The triple $(\Rkn,\varepsilon,()^{*})$, where $\varepsilon$ and $()^{*}$ are restricted to the substructures of the form $\Rkn\As$, is also a comonad.  

There is a comonad morphism $\nu:\Rk \rightarrow \Pk$ with components $\nu_{\As}:\Rk\As \rightarrow \Pk\As$, where $\nu_{\As}(s,i) = s[1,i]$, i.e.\ the length $i$ prefix of $s$.

\begin{prop}
  $\nu:\Rk \rightarrow \Pk$ is a comonad morphism.
\end{prop}  
\begin{proof}
We must confirm that $\nu$ is indeed a natural transformation, i.e.\ the following diagram commutes in $\Rsig$, for every $f:\As \rightarrow \Bs$: 
\begin{center}
  \begin{tikzcd}
    \Rk\As \ar[r, "\nu_{\As}"] \ar[d,"\Rk f"']  
    & \Pk \As \ar[d, "\Pk f"] \\
    \Rk\Bs \ar[r,"\nu_{\Bs}"] 
    & \Pk\Bs
  \end{tikzcd}
\end{center}
It is clear that this diagram commutes by observing that $\Rk f$ preserves the prefix relation on sequences. 
We present both comonads in the standard forms $(\Rk,\varepsilon,\delta)$ and $(\Pk,\varepsilon',\delta')$.
To confirm $\nu$ is a comonad morphism, we must show that the following diagrams commute in the category of endofunctors over $\Rsig$:
\begin{figure}[H]
  \centering
    \begin{tikzcd}
      \Rk \ar[r, "\nu"] \ar[rd, "\varepsilon"]
      &  \Pk \ar[d, "\varepsilon'"] \\
      & \textsf{Id}_{\Rsig} 
    \end{tikzcd}
 \;\; 
    \begin{tikzcd}
      \Rk^{2}  \ar[r, "\Rk \nu"] & \Rk\Pk \ar[r,  "\nu"] & \Pk^{2} \\
      \Rk \ar[u, "\delta"] \ar[rr, "\nu"]{} & & \Pk \ar[u, "\delta'"]
    \end{tikzcd}  
\end{figure}
The diagram on the left states that the last pebbled element of $\nu_{\As}(s,i) = s[1,i]$ is the $i$-th element pebbled in $s$ which is clear by definition.
To confirm the diagram on the right, recall that $\delta_{\As}$ and $\delta'_{\As}$ are the coextensions of $\id_{\Rk\As}$ and $\id_{\Pk\As}$ respectively. 
Explicitly, for $s = [(p_1,a_1),\dots,(p_n,a_n)]$ we have:
\begin{align*}
  \delta_{\As}(s,i) &= ([(p_1,(s,1)),\dots,(p_n,(s,n))],i); \\
  \delta'_{\As}(s) &= [(p_1,s[1,1]),\dots,(p_n,s[1,n])].
\end{align*}
The confirmation is then straightforward:
\begin{align*}
  &\nu_{\As} \circ \Rk\nu_{\As} \circ \delta_{\As}(s,i) \\ &=\nu_{\As} \circ \Rk\nu_{\As}([(p_1,(s,1)),\dots,(p_n,(s,n))],i)  & (1) \\
                                                       &= \nu_{\As} ([(p_1,\nu_{\As}(s,1)),\dots,(p_n,\nu_{\As}(s,n))],i) & (2) \\
                                                       &= \nu_{\As} ([(p_1,s[1,1]),\dots,(p_n,s[1,n])],i) & (3) \\
                                                       &= [(p_1,s[1,1]),\dots,(p_i,s[1,i])] & (4) \\
                                                       &= \delta'_{\As}(s[1,i]) &(5)\\
                                                       &= \delta'_{\As} \circ \nu_{\As}(s,i); & (6)
\end{align*}
where $(1)$ and $(5)$ follow by $\delta'_{\As}$ as above, $(3)$,$(4)$ and $(6)$ follow by the definition of $\nu_{\As}$ and $(2)$ follows by the functoriality of $\Rk$.
\end{proof}
\section{Coalgebras and Path Decompositions}
\label{sec:coalgebras}
We return to the initial motivation for constructing the pebble-relation comonad. 
That is, to give a categorical definition for the combinatorial parameter of pathwidth.
There are many different characterisations of pathwidth. 
The original definition, introduced by Robertson and Seymour in \cite{robertson1983}, is in terms of path decompositions of a structure $\As$.
We will show that coalgebras $\As \rightarrow \Rk \As$ correspond to path decompositions of $\As$ of width $< k$.
Given the relationship between $\Rk$ and $\Pk$, our proof is very similar to the one given in \cite{abramsky2017,abramskyResources2021} demonstrating that coalgebras over the pebbling comonad correspond to tree decompositions.
Intuitively, a sequence of pebbled elements with a common root forms a tree, so the mapping $f:A \rightarrow \Pk A$ associates with $a \in A$ a node in that tree. 
By contrast, a mapping $f:A \rightarrow \Rk A$ associates with an element $a \in A$ a full path of pebbled elements and index into that path. 
It is this intuition that explains why coalgebras of $\Pk$ correspond to tree decompositions, while coalgebras of $\Rk$ correspond to path decompositions.

\begin{defi}[path decomposition]
  Given a finite $\sigma$-structure $\As$, a \textit{path decomposition of $\As$} is a triple $(X,\leq_X,\lambda)$, where $(X,\leq_X)$ is a linearly ordered set and $\lambda:X \rightarrow \wp (A)$ is a function satisfying the following conditions:  \begin{enumerate}[label=(PD\arabic*),align=left] 
  \item For every $a \in A$, there exists $x \in X$ such that $a \in \lambda(x)$.\label{cond:pd0} 
  \item If $a \arc a' \in \As$, then $a,a' \in \lambda(x)$ for some $x \in X$. \label{cond:pd1}  
  \item For all $y \in [x,x']$, $\lambda(x) \cap \lambda(x') \subseteq \lambda(y)$, where $[x,x']$ is an interval with respect to $\leq_X$. \label{cond:pd2}
\end{enumerate}
\end{defi}
The \emph{width of a path decomposition} $(X,\leq_X,\lambda)$ for $\As$ is given by $k = \max_{x \in X} |\lambda(x)| - 1$. \label{cond:pdw} 
\begin{defi}[pathwidth]
The pathwidth of a $\sigma$-structure $\As$, denoted by $\pw(\As)$, is the least $k$ such that $\As$ has a path decomposition of width $k$.  
\end{defi}

In order to show the correspondence between tree decompositions of width $< k$ and coalgebras over $\Pk$, Abramsky et al.\ made use of an intermediate structure called a $k$-pebble forest cover\footnote{This is originally called $k$-traversal in \cite{abramsky2017}.} \cite{abramsky2017,abramskyResources2018}. 
We will use the analogous notion in the linear-ordered case -- the $k$-pebble linear forest cover.

\begin{defi}[$k$-pebble linear forest cover]
  Given a $\sigma$-structure $\As$, a \textit{$k$-pebble linear forest cover for $\As$} is a tuple $(\mathcal{F},p)$, where $\mathcal{F} = \{(S_i,\leq_i)\}_{i \in I}$ is a partition of $A$ into (possibly infinitely many) linearly ordered subsets and $p:A \rightarrow \setk$ is a pebbling function such that the following conditions hold:
\begin{enumerate}[label=(FC\arabic*),align=left]
  \item If $a \arc a' \in \As$, then there exists $i\in I$ such that $a,a' \in S_i$. \label{cond:fc1} 
  \item If $a \arc a' \in \As$ and $a \leq_i a'$, then for all $b \in (a,a']_{i} \subseteq S_i$, $p(b) \not= p(a)$. \label{cond:fc2}
  \end{enumerate}
\end{defi}
We aim to show that the existence of a path decomposition $(X,\leq_{X},\lambda)$ of width $< k$ for a finite $\As$ is equivalent to the existence of a $k$-pebble linear forest cover for $\As$. 
In order to do so, we need to define a pebbling function $p:A \rightarrow \setk$. 
We accomplish this by defining functions $\tau_{x}:\lambda(x) \rightarrow \setk$ on the subset associated with a node $x \in X$ in the path decomposition. 
If we define these functions in a consistent manner, then they can be `glued' together to obtain $p$. 
We use the definition below to identify consistent families of $\tau_{x}$.
\begin{defi}[$k$-pebbling section family]
  Given a path decomposition $(X,\leq_{X},\lambda)$ of width $< k$ for $\As$, we define a \textit{$k$-pebbling section family for $(X,\leq_{X},\lambda)$} as a family of functions $\{\tau_{x}:\lambda(x) \rightarrow \setk\}$ indexed by $x \in X$, such that the following conditions hold:
  \begin{enumerate}
    \item (Locally-injective) For every $x \in X$, $\tau_{x}$ is an injective function.
    \item (Glueability) For every $x,x' \in X$, \[\tau_{x}|_{\lambda(x) \cap \lambda(x')} = \tau_{x'}|_{\lambda(x) \cap \lambda(x')}.\]
  \end{enumerate}
\end{defi}
We show that every path decomposition has a pebbling section family. 
\begin{lem}\label{lem:kPebbleScheme}
  If $(X,\leq_{X},\lambda)$ is a path decomposition of width $< k$, then $(X,\leq_{X},\lambda)$ has a $k$-pebbling section family $\{\tau_{x}\}_{x \in X}$. 
\end{lem}\begin{proof}
  By induction on the linear order $\leq_{X}$ for every $x \in X$, we construct a $k$-pebbling section family $\{\tau_{z}\}_{z \in {\downarrow x}}$ for ${(Y, \leq_{Y}, \lambda|_{Y})}$,
  where $Y = {\darr x}$ and $\leq_{Y} = {\leq_{X}} \cap (Y \times Y)$.
  
  \textit{Base Case:} Suppose $r$ is the $\leq_{X}$-least element. 
  By definition \ref{cond:pdw}, the cardinality of $\lambda(r)$ is $\leq k$, thus we can enumerate the elements via an injective function $\tau_{r}:\lambda(r) \rightarrow \setk$. 
  By construction, $\tau_{r}$ is injective, so $\{\tau_{r}\}$ is locally-injective. 
  Glueability follows trivially as every $x \in {\downarrow r}$ is equal to $r$.
  
  \textit{Inductive Step:} Let $x'$ be the immediate $\leq_{X}$-successor of $x$.
  By the induction hypothesis, there exists a $k$-pebbling section family $\{\tau_{y}\}_{y \in {\downarrow x}}$. 
  Let $V_{x'}$ denote the subset of `new' elements $a \in \lambda(x')$ such that $a \not\in \lambda(y)$ for every $y <_X x'$.

  \emph{Claim:} For every $y <_X x'$, $\lambda(y) \cap \lambda(x') \subseteq \lambda(x) \cap \lambda(x')$. 
  We show that this claim holds: By \ref{cond:pd2}, for all $z \in [y,x']$, $\lambda(y) \cap \lambda(x') \subseteq \lambda(z)$. 
  In particular, since $y \leq_X x <_X x'$, $\lambda(y) \cap \lambda(x') \subseteq \lambda(x)$. 
  Therefore,  $\lambda(y) \cap \lambda(x') \subseteq \lambda(x) \cap \lambda(x')$. 
  
  From the claim and the definition of $V_{x'}$, we have that $\lambda(x') = (\lambda(x) \cap \lambda(x')) \sqcup V_{x'}$, where $\sqcup$ denotes a disjoint union. 
  This allows us to define $\tau_{x'}:\lambda(x') \rightarrow \setk$ by cases on each of these parts. 
  Fix an injective function $\upsilon_{x'}:V_{x'} \rightarrow \setk$ enumerating $V_{x'}$. 
  Let $\{i_1,\dots,i_m\}$ enumerate the elements of $\setk$ not in the image of $\tau_{x}|_{\lambda(x) \cap \lambda(x')}$.
  Define $\tau_{x'}:\lambda(x') \rightarrow \setk$ as
      \begin{equation*}
        \tau_{x'}(a) = \begin{cases}
          \tau_{x}(a) & \text{if $a \in \lambda(x) \cap \lambda(x')$}; \\
        i_j & \text{if $a \in V_{x'}$ and $\upsilon_{x'}(a) = j$}.
      \end{cases}
      \end{equation*}
      Injectivity of $\tau_{x'}$ follows from the injectivity of $\tau_{x}|_{\lambda(x) \cap \lambda(x')}$ and $\upsilon_{x'}$. 
      To verify glueability, it suffices to check that $\tau_{x'}|_{\lambda(y) \cap \lambda(x')} = \tau_{y}|_{\lambda(y) \cap \lambda(x')}$ for $y \in {\downarrow x'}$. 
      Since $\{\tau_{y}\}_{y \in {\downarrow x}}$ is a $k$-pebbling section family, for all $y \in {\downarrow x}$, $\tau_{y}|_{\lambda(y) \cap \lambda(x)} = \tau_{x}|_{\lambda(y) \cap \lambda(x)}$.
      By construction, $\tau_{x}|_{\lambda(x) \cap \lambda(x')} = \tau_{x'}|_{\lambda(x) \cap \lambda(x')}$. 
      By the claim, we have that $\lambda(y) \cap \lambda(x') \subseteq \lambda(x) \cap \lambda(x')$. 
      Therefore, $\tau_{y}|_{\lambda(y) \cap \lambda(x')} = \tau_{x'}|_{\lambda(y) \cap \lambda(x')}$.
\end{proof}
This leads to the following two theorems that are essential to the categorical characterisation of pathwidth.
\begin{thm}
The following are equivalent for all finite $\sigma$-structures $\As$:
  \begin{enumerate}
    \item $\As$ has a path decomposition of width $< k$.
    \item $\As$ has a $k$-pebble linear forest cover.
  \end{enumerate}
  \label{thm:pathToCover}
\end{thm}
\begin{proof}
    ${(1) \Rightarrow (2)}$ Suppose $(X,\leq_{X},\lambda)$ is a path decomposition of $\As$ of width $< k$. 
    We define a family of linearly ordered sets $\{(S_i,\leq_i)\}$, where each $S_i$ is the vertex set of a connected component of $\Gf(\As)$. 
    To define the order $\leq_{i}$, we define an order on $\leq_{A}$ and realise $\leq_{i}$ as the restriction of $\leq_{A}$ to $S_i$. 
    For every $a \in A$, let $x_{a} \in X$ denote the $\leq_{X}$-least element in $X$ such that $a \in \lambda(x_a)$. 
    Such an $x_a$ always exists by \ref{cond:pd0}.  
    By lemma \ref{lem:kPebbleScheme}, there exists a $k$-pebbling section family $\{\tau_{x}\}_{x \in X}$. 
    We then define $\leq_{A}$ as follows: 
    \[a \leq_{A} a' \Leftrightarrow x_{a} <_{X} x_{a'} \text{ or } \tau_{x}(a) \leq \tau_{x}(a') \text{ if $x_a = x_{a'} = x$}.\]
    
    The glueability condition on $k$-pebble section family $\{\tau_x\}$ allows us to obtain a well-defined pebbling function $p:A \rightarrow \setk$ from $\tau_x$.
    Explicitly, thinking of functions as their sets of ordered pairs, $p = \bigcup_{x \in X} \tau_{x}$.
    The tuple $(\{(S_i,\leq_i)\},p)$ is a $k$-pebble linear forest cover.
    
    To verify that $\{(S_i,\leq_i)\}$ is a partition of $A$ into linearly ordered subsets, we observe that by construction each $S_i$ is a connected component of $A$, and so $\{S_i\}$ partitions $A$. 
    Suppose $a,a' \in S_i$, then by $\leq_{X}$ being a linear order, either $x_{a} <_X x_{a'}$, $x_{a} >_X x_{a'}$, or $x_{a} = x_{a'}$. 
    If $x_{a} <_X x_{a'}$ or $x_{a} >_X x_{a'}$, then $a <_{i} a'$ or $a >_{i} a'$ by the definition of $\leq_{i}$. 
    If $x_{a} = x_{a'} = x$, then either $\tau_{x}(a) \leq \tau_{x}(a')$ or $\tau_{x}(a) \geq \tau_x(a')$ by the linear ordering $\leq$ on $\setk$. 
    Hence, in either case, $a \leq_{i} a'$ or $a \geq_{i} a'$, so $\leq_{i}$ is a linear ordering.
    
    To verify \ref{cond:fc1}, suppose $a \arc a' \in \As$. 
    This means $a,a'$ are connected in $\Gf(\As)$, and so are in the same connected component $S_i$ of $\Gf(\As)$.
    
    To verify \ref{cond:fc2}, suppose $a \arc a' \in S_i$, $a \leq_i a'$, and $b \in (a,a']_i$. 
    By definition of~${\leq_{i}}$, $x_{a} \leq_{X} x_{b} \leq_{X} x_{a'}$.
    Since $a \arc a'$, by~\ref{cond:pd1} there exists~$x \in X$ such that $a,a' \in \lambda(x)$. 
    By the definition of $x_{a'}$ as the $\leq_{X}$-least element of $X$ containing $a'$, we have $x_{a'} \leq_{X} x$.
    By transitivity of $\leq_X$ and $x_{a} \leq_X x_{a'} \leq_X x$, we have that $x_{a} \leq_X x$. 
    By \ref{cond:pd2}, for every $y \in [x_{a},x]_{X}$, $\lambda(x_a) \cap \lambda(x) \subseteq \lambda(y)$. 
    In particular, for $x_{b} \in [x_{a},x_{a'}]_{X} \subseteq [x_{a},x]_{X}$, we have $a \in \lambda(x_{b})$. 
    Hence, $a,b \in \lambda(x_{b})$ and by the injectivity of $\tau_{x_{b}}$, $\tau_{x_{b}}(a) \not= \tau_{x_{b}}(b)$.
    It follows that $p(a) \not= p(b)$.

    ${(2) \Rightarrow (1)}$ Suppose $\As$ has $k$-pebble linear forest cover given by the partition $\{(S_i,\leq_i)\}_{i \in \setn}$ and pebbling function $p:A \rightarrow \setk$.
    We define a linearly ordered set $(A,\leq_{A})$, where $\leq_{A}$ is the ordered sum of the family $\{(S_i,\leq_i)\}_{i \in \setn}$. 
    Explicitly, $a \leq_{A} a'$ iff $a \in S_i$, $a' \in S_j$ for $i < j$ or $a \leq_i a'$ for $i = j$. 
    We say that an element $a$ is an \textit{active predecessor} of $a'$ if $a \leq_{A} a'$ and for all $b \in (a,a']_{A}$, $p(b) \not= p(a)$. 
    Let $\lambda(a)$ be the set of active predecessors of $a$.
    The triple $(A,\leq_{A},\lambda)$ is a path decomposition of $\As$ of width $< k$. 
    
    To verify \ref{cond:pd0}, observe that, for every $a \in A$, $a$ is an active predecessor of itself since $a \leq_{A} a$ and $(a,a]_{A} = \varnothing$. 
    Hence, $a \in \lambda(a)$.

    To verify \ref{cond:pd1}, suppose $a \arc a' \in \As$. 
    By \ref{cond:fc1}, there exists an $S_i$ where $a,a' \in S_i$. 
    Without loss of generality, assume $a \leq_{i} a'$. 
    By \ref{cond:fc2}, for all $b \in (a,a']_{i}$, $p(b) \not= p(a)$. 
    Therefore, $a$ is an active predecessor of $a'$, so $a,a' \in \lambda(a')$.
    
    To verify \ref{cond:pd2}, suppose $b \in [a,a']_{A}$ and that ${c \in \lambda(a) \cap \lambda(a')}$. 
    By $c \in \lambda(a)$ and $b \in [a,a']_{A}$, we have that $c \leq_{A} a$ and $a \leq_{A} b$, so $c \leq_{A} b$. 
    By $c \in \lambda(a')$, for all $d \in (c,a']_{A}$, $p(c) \not= p(d)$. 
    In particular, for all $d \in (c,b]_{A}$, $p(c) \not= p(d)$.
    By definition, $c$ is an active predecessor of $b$, so $c \in \lambda(b)$.  
    
    To verify the width of the decomposition $< k$, we need to show that for every $a' \in A$, $|\lambda(a')| \leq k$.
    Assume for contradiction that $|\lambda(a')| > k$ for some $a' \in A$. 
    Consider the pebbling function restricted to $\lambda(a')$, $p|_{\lambda(a')}:\lambda(a') \rightarrow \setk$. 
    By the Pigeonhole Principle, there must exist $a,c \in \lambda(a')$ with $a \not= c$, such that $p(a) = p(c)$.
    Without loss of generality assume that $a <_{A} c$. 
    Since $a \in \lambda(a')$, $a$ is an active predecessor of $a'$, i.e.\ for all $b \in (a,a']_{A}$, $p(b) \not= p(a)$.
    In particular, since $c \in (a,a']_{A}$, as $a <_{A} c$ and $c \in \lambda(a')$, then $p(c) \not= p(a)$. This yields a contradiction.
    \end{proof}
In analogy to $\Pk$, where $k$-pebble forest covers correspond to $\Pk$-coalgebras, we also show that $k$-pebble linear forest covers correspond to $\Rk$-coalgebras. 
\begin{thm}\label{thm:coverToCoalgebra}
For all $\sigma$-structures $\As$, there is a bijective correspondence between the following:
\begin{enumerate}
    \item $k$-pebble linear forest covers of $\As$.
    \item coalgebras $\alpha:\As \rightarrow \Rk \As$.
\end{enumerate}
\end{thm}
  \begin{proof}
  ${(1) \Rightarrow (2)}$ Suppose $\As$ has $k$-pebble linear forest cover given by the partition into linear orders ${\{(S_i,\leq_i)\}_{i \in \setn}}$ and pebbling function $p:A \rightarrow \setk$.
  Since each $(S_i,\leq_i)$ is linearly ordered, we can present $(S_i,\leq_i)$ as a chain
  $a_1 \leq_{i} \dots \leq_{i} a_{m_i}$.
  We define
  \[t_i = [(p(a_1), a_1),\dots,(p(a_m), a_{m_i})].\]
  Intuitively, $t_i$ is the enumeration induced by the linear order $\leq_i$ of $S_i$ zipped with its image under $p$. 
  For every $a_j \in S_i$ let $\alpha_i:S_i \rightarrow \Rk\As$ be defined as $\alpha_i(a_j) = (t_i,j)$. 
  Let $\alpha:A \rightarrow \Rk A$ be $\alpha = \bigcup_{i \in \setn} \alpha_i$. 
  Since the collection of $S_i$ partitions $A$, $\alpha$ is well-defined. 
  We must show that the function $\alpha$ is a coalgebra $\alpha:\As \rightarrow \Rk \As$. 
  
  To verify that $\alpha$ is indeed a homomorphism, suppose $R \in \sigma$ is an $m$-ary relation and $R^{\As}(a_1,\dots,a_m)$.
  By \ref{cond:fc1}, $\{a_1,\dots,a_m\} \subseteq S_i$ for some $i \in \setn$. 
  Therefore, for all $j \in \setm$, $\alpha(a_j) = (t_i,z_j)$ for some $z_j \in \{1,\dots,|t_i|\}$. 
  Let $z$ be the maximal index amongst the $z_j$. 
  Assume $\alpha(a) = (t,z)$ for $a \in \{a_1,\dots,a_m\}$. 
  By \ref{cond:fc2}, for every $a_j$ and $b \in (a_j,a]_{i}$, $p(a_j) \not= p(b)$. 
  Therefore, $\pi_{A}(t_i,z_j)$ does not appear in $t_i(z_j,z]$.
  Hence, by the definition of $R^{\Rk \As}$ and the supposition that $R^{\As}(a_1,\dots,a_m)$, we obtain $R^{\Rk \As}((t_i,z_{1}),\dots,(t_i,z_{m}))$.

  To verify that $\alpha$ satisfies the counit-coalgebra law, suppose $a \in \As$. Then, by $\{S_i\}$ partitioning $A$, $a \in S_i$ for some $i \in \setn$. 
  Suppose $a$ is the $j$-th element in the $\leq_i$ linear ordering:
  \begin{align*}
    \varepsilon_{\As} \circ \alpha(a) &= \varepsilon_{\As} \circ \alpha_i(a)  \tag{\text{by supposition $a \in S_i$}} \\
                                   &= \varepsilon_{\As}(t_i,j) \tag{\text{def. $\alpha_i$}} \\
              &= a \tag{\text{def. $t_i$, $\varepsilon_A$}}.
  \end{align*}
  To verify that $\alpha$ satisfies the comultiplication-coalgebra law observe that  
  \begin{align*}
    &\delta_{\As} \circ \alpha(a) 
    \\&= \delta_{\As} \circ \alpha_{i}(a)  \tag{\text{by supposition $a \in S_i$}} \\
                                 &= \delta_{\As}(t_i,j)  \tag{\text{def.  $\alpha_i$}} \\
                                 &= ([(p(a_1),(t_i,1)),\dots,(p(a_{m_i}),(t_i,{m_i}))],j) \tag{\text{def. $t_i$, $\delta_{\As}$}} \\
                                 &= ([(p(a_1),\alpha(a_1)),\dots,(p(a_{m_i}),\alpha(a_{m_i}))],j)  \tag{\text{def.  $\alpha$}} \\
                                 &= \Rk\alpha(t_i,j)  \tag{\text{def.  $t_i$ and functoriality of $\Rk$}} \\
                                 &= \Rk\alpha \circ \alpha(a)  \tag{\text{def. $\alpha$}}.
  \end{align*}

  ${(2) \Rightarrow (1)}$. We define a family of linearly ordered subsets $\{(S_t,\leq_t)\}$ of $A$:
  \begin{align*}
    S_{t} &:= \{a \mid \alpha(a) = (t,j) \text{ for some $j \in [|t|]$}\}; \\
    a \leq_t a' &\Leftrightarrow \alpha(a) = (t,j)\text{, } \alpha(a') = (t,j') \text{ and } j \leq j'. 
  \end{align*}
  By the counit-coalgebra axiom, if $\alpha(a) = (t,j)$, then $a$ is the $j$-th element in the chain $S_{t}$.
  By the comultiplication-coalgebra axiom, the relation $\leq_{t}$ is a linear order. 
  Let $p:A \rightarrow \setk$ be $p = \pi_{A} \circ \alpha$. 
  The tuple $(\{(S_t,\leq_t)\},p)$ is a $k$-pebble linear forest cover of $\As$. 

  To verify \ref{cond:fc1}, suppose that ${a \arc a'}$ and that ${\alpha(a) = (t,j)}$ and ${\alpha(a') = (t',j')}$. 
  By $\alpha$ being a homomorphism, ${\alpha(a) \arc \alpha(a')}$, so ${(t,j) \arc (t',j')}$. 
  However, by the definition of $R^{\Rk \As}$, for all $R \in \sigma$, elements of $\Rk\As$ are only related if they are part of the same pebble play, so $t = t'$. 
  By definition, $a,a' \in S_t$. 

  To verify \ref{cond:fc2}, suppose that $a \arc a'$ with $a \leq_t a'$ and $b \in (a,a']_{t}$. 
  We want to show that $p(b) \not= p(a)$. 
  Since ${a \arc a'}$, then there exists some $m$-tuple $\vec{a} \in R^{\As}$ for some $m$-ary relation $R \in \sigma$.
  By $\alpha$ being a homomorphism, there exist $(t,j), (t,j') \in \alpha(\vec{a}) \in R^{\Rk\As}$ for some $j \leq j' \in [|t|]$ such that $\alpha(a) = (t,j)$ and $\alpha(a') = (t,j')$.
  Moreover, since $b \in (a,a']_{t}$, then there exists $i \in (j,j']$ such that $\alpha(b) = (t,i)$. 
  By construction, ${p(b) = \pi_{A} \circ \alpha(b) = \pi_{A}(t,i)}$.
  However, by the first condition in the definition of $R^{\Rk \As}$, $\pi_{A}(t,j)$ does not appear in $t(j,j']$, so ${p(b) = \pi_{A}(t,i) \not= \pi_{A}(t,j) = p(a)}$. 
\end{proof}
We can now use the coalgebra number $\kappa^{\PR}$ to define pathwidth.

\begin{cor}\label{cor:pwd}
For all finite $\sigma$-structures $\As$, $\pw(\As) = \kappa^{\PR}(\As) - 1$.
\end{cor}
\begin{proof}
By Theorem \ref{thm:pathToCover} and Theorem \ref{thm:coverToCoalgebra}, a structure $\As$ has a path decomposition of width $< k$ iff $\As$ has a coalgebra $\As \rightarrow \Rk \As$.
Hence, $\pw(\As) + 1 \leq \kappa^{\PR}(\As)$ by the definition of $\pw(\As)$ as the minimal width of a path decomposition for $\As$, and $\kappa^{\PR}(\As) \leq \pw(\As) + 1$ by $\kappa^{\PR}$ being the minimal index for a $\Rk$-coalgebra of $\As$.  
\end{proof}
We can extend Theorem \ref{thm:coverToCoalgebra} to capture an equivalence between two categories. 
Consider the category $\TsigRk$ of forest-ordered $\sigma$-structures $(\As,\leq_A,p)$ with $\As \in \Rsig_0$ and pebbling function $p:A \rightarrow \setk$ satisfying the following conditions:
\begin{enumerate}[label=(E),align=left]
    \item If $a \arc a' \in \As$, then either $a \leq_{A} a'$ or $a' \leq_{A} a$.
\end{enumerate}
\begin{enumerate}[label=(P),align=left]
    \item If $a \arc a' \in \As$ and $a \leq_{A} a'$, then for all $b \in (a,a']_{A}$, $p(b) \not= p(a)$.
\end{enumerate}
\begin{enumerate}[label=(L),align=left]
    \item  For every $a \in A$, $\uarr a$ is linearly-ordered by $\leq_{A}$. 
\end{enumerate}
The condition (L), taken together with $\leq_{A}$ being a tree-order, i.e.\ for every $a \in A$, ${\darr a}$ is linearly ordered by $\leq_{A}$, means that $(A,\leq_{A})$ is a disjoint union of linearly ordered sets. 
In other words, $(A,\leq_{A})$ is a linear forest. 
Morphisms in $\TsigRk$ are $\sigma$-morphisms that preserve paths of $\leq$ and pebbling functions. 
There is an evident forgetful functor $U_k:\TsigRk \rightarrow \Rsig$, which yields that $(\As,\leq_A,p) \mapsto \As$. 

Consider the functor $F_k:\Rsig \rightarrow \TsigRk$ with object mapping $\As \mapsto (\Rk\As,\leq^{*},\pi_{\As})$, where $(t,i) \leq^{*} (t',j)$ iff $t = t'$ and $i \leq j$, and morphism mapping $f \mapsto \Rk f$. The functor $F_k$ is right adjoint to $U_k$. In fact, this adjunction yields an equivalence between $\TsigRk$ and $\EM(\Rk)$. 
\begin{thm}
For each $k > 0$, $U_k \dashv F_k$ is a comonadic adjunction. Moreover, $\Rk$ is the comonad arising from this adjunction.
\end{thm}
\begin{proof}
By theorem \ref{thm:coverToCoalgebra}, the objects of the two categories are in bijective correspondence. 
Thus to complete the argument, we must show that a coalgebra morphism coincides with a $\TsigRk$-morphism. 
Coalgebra morphisms preserve paths and pebble indices, and this is equivalent to the conditions for a $\TsigRk$-morphism. 
\end{proof}
\section{Morphism and Isomorphism Power Theorems}
\label{sec:equivalences}
While the construction of $\Rk$ is motivated by capturing pathwidth as coalgebras, the other Spoiler-Duplicator game comonads $\CC_k$, explored in previous work \cite{abramsky2017, abramskyResources2018, oconghaile2020, abramsky2020}, were constructed by capturing Duplicator's winning strategies as coKleisli morphisms of $\CC_k$. 
In this section we prove a morphism power theorem for $\Rk$, showing that coKleisli morphisms $\Rk\As \rightarrow \Bs$ correspond to Duplicator's winning strategies in a one-sided pebble-relation game. 
This game characterises preservation of sentences in the existential positive $k$-variable logic with restricted conjunctions $\pRLogicK$. 
We then prove an isomorphism power theorem for $\Rk$ which works analogously by showing that coKleisli isomorphisms $\Rk\As \rightarrow \Bs$ correspond to Duplicator's winning strategies in a new type of pebble game which will be described in detail. We conclude the section with an ensuing corollary of the exhibited results. 

\subsection{Morphism power theorem}

We begin by introducing Dalmau's $k$-pebble relation game on $\As$ and $\Bs$, denoted by $\exists^{+} \PebR_k(\As,\Bs)$, which characterises the relation $\As \Rrightarrow^{\pRLogicK} \Bs$.
\begin{defi}[Dalmau's $k$-pebble relation game \cite{dalmau2005}]
Each round of the game ends with a pair $(I,T)$ where $I \subseteq A$ is a domain such that $|I| \leq k$ and $T \subseteq \Rsig_1(\As|_{I},\Bs)$ is a set of $\sigma$-morphisms from $\As|_{I}$ to $\Bs$.
At round $0$, $I = \varnothing$ and $T = \{\lambda\}$, where $\lambda$ is the unique function from $\varnothing$ to $\Bs$. 
At each subsequent round $n > 0$, let $(I,T)$ denote the configuration of the previous round $n-1$.
Spoiler then chooses between two possible moves: 
    \begin{enumerate}
        \item A shrinking move, where Spoiler chooses a smaller domain $I' \subseteq I$.
        \begin{itemize}
            \item Duplicator then chooses $T'$ to be the set of restrictions of the morphisms in $T$ to $I'$, i.e.\ $T' = \{h|_{I'} \mid h \in T\}$.
        \end{itemize}
        \item A blowing move when $|I| < k$, where Spoiler chooses a larger domain $I' \supseteq I$ with $|I'| \leq k$. 
        \begin{itemize}
            \item Duplicator then chooses a set of $\sigma$-morphisms $T'$ from $\As|_{I'}$ to $\Bs$ such that $T'|_I  = \{h|_{I} \mid h \in T'\} \subseteq T$.
        \end{itemize}
    \end{enumerate}
At the end of the $n$-th round, the configuration is $(I',T')$.
Duplicator wins the $n$-th round if $T'$ is non-empty. 
Duplicator wins the game if she can play forever. Otherwise, Spoiler wins the game.
\end{defi}

Intuitively, $\exists^{+} \PebR_k$ is a revised version of the $k$-pebble game in which Duplicator is given the advantage of non-determinism.
That is, she can respond with a set of partial homomorphisms.
At each round she is only obligated to extend some of the partial homomorphisms chosen in the previous round. 
In fact, we can recover the one-sided $k$-pebble game $\exists^{+} \Peb_k$ by insisting that Duplicator always responds with a singleton set $T$. 

In the same paper, Dalmau proves the following result:
\begin{thmC}[\cite{dalmau2005}]\label{dalmau2005result}
The following are equivalent for all $\sigma$-structures $\As$ and $\Bs$:
\begin{enumerate}
    \item Duplicator has a winning strategy in $\exists^{+}\PebR_k(\As,\Bs)$.
	\item $\As \Rrightarrow^{\pRLogicK}\Bs$.
	\end{enumerate}	
\end{thmC}

In fact, Dalmau shows that if Duplicator has a winning strategy, then Duplicator has a default winning strategy, which he calls complete, by playing the set of all partial morphisms $T$ on Spoiler's choice of domain $I$. 

We will show that coKleisli morphisms $\Rk\As \rightarrow \Bs$ capture Duplicator's winning strategies in $\exists^{+} \PebR_k(\As,\Bs)$.
Note that it is not clear in what way elements of $\Rk\As$ correspond to Spoiler's plays in $\exists^{+} \PebR_k$ on $\As$. 
It is also not clear in what way functions $\Rk\As \rightarrow \Bs$ correspond to Duplicator's responses in $\exists^{+} \PebR_k(\As,\Bs)$.
Fortunately, we can view elements of $\Rk\As$ as Spoiler's plays in a different, yet equivalent (Theorem \ref{equivgame}), game -- the all-in-one  $k$-pebble game $\exists^{+} \PPeb_k(\As,\Bs)$.\footnote{This game was discovered independently by the authors. The authors later discovered that a similar game appeared in a currently unavailable draft written by Iain A. Stewart.}
\begin{defi}[all-in-one $k$-pebble game]
The game is played in one round during which:
\begin{itemize}
    \item Spoiler chooses a sequence $s = [(p_1,a_1),\dots,(p_n,a_n)]$ of pebble placements on elements of the $\sigma$-structure $\As$.
    \item Duplicator responds with a compatible (same length and corresponding pebble at each index) sequence of pebble placements $t = [(p_1,b_1),\dots,(p_n,b_n)]$ on elements of $\Bs$.
\end{itemize}
Duplicator wins the game if for every index $i \in \setn$, the relation 
\[\gamma_i = \{(\last_{p}(s[1,i]),\last_{p}(t[1,i])) \mid p \in \pebbles(s[1,i]) \}\] 
is a partial homomorphism from $\As$ to $\Bs$. 
\end{defi}
\begin{exa}[separating the pebble games]
In the $2$-pebble game on $\As$ and $\Bs$ below, Spoiler has a winning strategy. 
Let $R$ and $G$ denote the unary predicates `red' and `green' respectively.
The winning strategy of Spoiler is expressed by the $\exists^+\mathscr{L}^2$-formula
\begin{align*}
		\exists x\exists y\Big( &\exists y( Exy\wedge R y)\wedge \exists y( Exy\wedge Gy)\wedge \\& \exists x(Eyx\wedge Rx)\wedge \exists x(Eyx\wedge Gx)\wedge Exy\Big),
	\end{align*}
which is true in $\As$ but not in $\Bs$.
In contrast, Duplicator wins the all-in-one $2$-pebble game on $\As$ and $\Bs$.
This is straightforward, since the only difference between the two structures (that is expressible in $\exists^+\mathscr{L}^2$) is the property characterised by the formula above, which cannot be converted to a formula in $\exists^+{\curlywedge}\mathscr{L}^2$. 

\begin{figure}[H]
    \centering
    \includegraphics[scale=0.2]{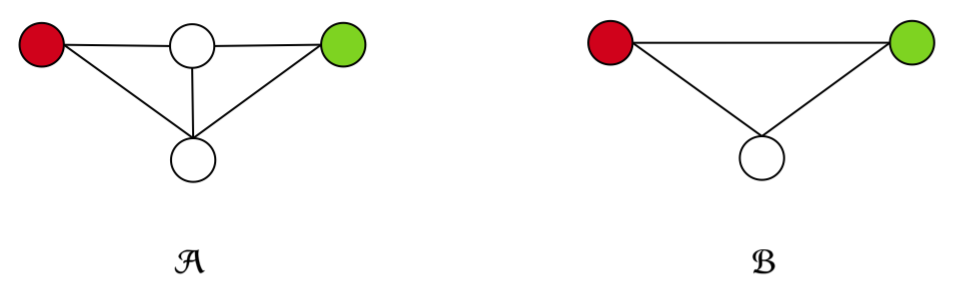}
    \caption{Separating example between the $k$-pebble game and the all-in-one $k$-pebble game for $k=2$.}
    \label{fig:my_label}
\end{figure}
\end{exa}

Besides its intuitive appeal and similarity to the original pebble game, as opposed to Dalmau's version, the all-in-one $k$-pebble game is a `real' game in the sense that Duplicator has a choice and there is no default play that ensures a victory whenever one exists.
Moreover, the elements of the comonad $\Rk$ are plays in the all-in-one version of the game rather than the pebble-relation game, hence establishing this connection contributes to developing a more coherent relationship between the game and the logic.
As we will see in the next subsection, this version of the game is also easier to generalise to the bijective case and work with in that setting. 

In order to ensure that we capture $k$-variable logic with equality, we must impose an additional $I$-morphism condition on morphisms of type $\Rk\As \rightarrow \Bs$. 
We can formulate the $I$-morphism condition by using the machinery of relative comonads.  
In order to do so, we expand the signature from $\sigma$ to $\sigma^{+} = \sigma \cup \{I\}$ which includes a new binary relation $I$, and we consider the functor $J:\Rsig \rightarrow \Rsigplus$, where a {$\sigma$-structure} $\As$ is mapped to the same $\sigma$-structure with $I^{J\As}$ interpreted as the identity relation, i.e.\ $I^{J\As} = \{(a,a) \mid a \in \As\}$. 
This leads us to the following theorem:

\begin{thm}\label{equivgame}
The following are equivalent for all {$\sigma$-structures} $\As$ and $\Bs$:
\begin{enumerate}
    \item Duplicator has a winning strategy in $\exists^{+} \PPeb_k(\As,\Bs)$.
    \item There exists a coKleisli morphism $f:\Rk\As \rightarrow \Bs$.
    \item There exists a coKleisli morphism $f^{+}:\Rk J \As \rightarrow J\Bs$.
    \item Duplicator has a winning strategy in $\exists^{+} \PebR_k(\As,\Bs)$.
\end{enumerate}

\end{thm}
\begin{proof}
${(1) \Rightarrow (2)}$ We first define the coextension $\sigma$-morphism $f^{*}:\Rk\As \rightarrow \Rk\Bs$ and note that $f = \varepsilon_{\Bs} \circ f^{*}$. 
Suppose Duplicator has a winning strategy in $\exists^+ \PPeb_k(\As,\Bs)$. 
Then for every sequence of pebble placements $s = [(p_1,a_1),\dots,(p_n,a_n)]$ on $\As$, Duplicator responds with a sequence of pebble placements $t = [(p_1,b_1),\dots,(p_n,b_n)]$ with the same length and corresponding pebble placement at each index.
We define $f^{*}(s,i) = (t,i)$ for every $i \in \setn$. 

To verify that $f^{*}$ is indeed a homomorphism, suppose that $R^{\Rk\As}((s,i_1),\dots,(s,i_m))$ and let $i = \max\{i_1,\dots,i_m\}$.
By the active pebble condition, for all $i_j \in \{i_1,\dots,i_m\}$, the pebble appearing at $i_j$ does not appear in $s(i_j,i]$. 
Hence, the element $\varepsilon_{\As}(s,i_j)$ was the last element pebbled by the pebble $\pi_{A}(s,i_j)$ in $s[1,i]$, and so $\varepsilon_{\As}(s,i_j)$ is in the domain of $\gamma_i$. 
By $\gamma_i$ being a partial homomorphism, $R^{\Bs}(b_{i_1},\dots,b_{i_m})$, thus $R^{\Bs}(\varepsilon_{\Bs}(t,i_1),\dots,\varepsilon_{\Bs}(t,i_m))$.
The active pebble condition follows from $t$ being compatible with $s$, so $R^{\Rk\Bs}((t,i_1),\dots,(t,i_m))$. 

${(2) \Rightarrow (3)}$ We say that a sequence ${s = [(p_1,a_1),\dots,(p_n,a_n)]}$ is \textit{duplicating} if there exists $i < j \in \setn$ such that $\pi_{A}(s,i)$ does not appear in $s(i,j]$ (i.e.\ $p_i$ is active at index $j$) with $a_i = a_j$.
Let $s'$ denote the longest subsequence of $s$ such that $s'$ is non-duplicating and for every $j \in \setn$ there exists $j' \leq j$ with $\varepsilon_{\As}(s',j') = \varepsilon_{\As}(s,j)$. 
Such an $s'$ can be shown to always exist for every sequence of pebble placements $s$. 
This can be shown inductively by removing moves $(p_j,a_j)$ that are duplicating the placement of a different active pebble $p_i \not= p_j$ on the same element $a_i = a_j$. 
We define $f^{+}(s,j) = f(s',j')$.

To verify that $f^{+}:\Rk J \As \rightarrow J \Bs$ is a $\sigma^{+}$-morphism, suppose $((s,i),(t,j) \in I^{\Rk J \As}$. 
By the equality condition, $s = t$. 
If $s$ is not duplicating, then $i = j$ and so $f^{+}(s,i) = f^{+}(t,j)$; therefore $(f^{+}(s,i),f^{+}(t,j)) \in I^{J\Bs}$ by $I^{J\Bs}$ being equality on $\Bs$. 
If $s$ is duplicating, then by construction there exists some non-duplicating $s'$ such that $f^{+}(s,i) = f^{+}(t,j) = f^{+}(s',j')$. 
As in the previous case, since $f^{+}(s,i) = f^{+}(t,j)$, we have that $(f^{+}(s,i),f^{+}(t,j)) \in I^{J\Bs}$. 
For the other relations $R \in \sigma \subseteq \sigma^{+}$, $f^{+}$ preserves $R$ because $f$ preserves $R$.

${(3) \Rightarrow (4)}$ For each round $n$ of the game with Spoiler choosing the domain $I$, we can construct a (possibly empty) sequence of pebbled elements $s$ such that $I \subseteq \Active(s)$ inductively. 
Throughout the induction, we will keep track of the active pebbles $P = \Active(s)$ in $s$.
For $n = 0$, $I = \varnothing$, so $s$ is the empty sequence and $P = \varnothing$.
For the inductive step, we assume for the inductive hypothesis, that $s$ was constructed in round $n-1$ from the move $I$ and $P = \Active(s)$. 
There are two cases for Spoiler's move. If $I' \subseteq I$ is a shrinking move, then $s' = s$ 
and 
\[ P' = P \backslash \{p \mid \last_{p}(s) \in I \backslash I'\}. \]
If $I' = I \cup \{a_1,\dots,a_j\}$ is growing move for $j \leq k$, then let
\[ s' = s[(p_1,a_1),\dots,(p_j,a_j)] \] 
for $p_1,\dots,p_j \not\in P$. 
Clearly, $P' = \Active(s) = P \cup \{p_1,\dots,p_j\}$. 

We construct Duplicator's response in round $n$ by considering the set 
    \[Z = \{t[1,m] \mid f^{\dagger}(s',m) = (t,m) \text{ and } s'[1,m] = s \},\]
where the map $f^{\dagger}:\Rk J \As \rightarrow \Rk J \Bs$ is the coextension of $f^{+}:\Rk J \As \rightarrow J \Bs$. 
For each $u = t[1,m] \in Z$, we can form the relation $\gamma_{s,u} = \{(\last_{p}(s),\last_{p}(u)) \mid p \in \pebbles(s)\}$ of pairs of active pebbled elements.
Since $f^{\dagger}$ is a $\sigma^{+}$-morphism, each $\gamma_{s,u}$ is a partial homomorphism. 
Hence, Duplicator's response $T = \{\gamma_{s,u} \mid u \in Z\}$ is a winning move in $\exists^{+}\PebR_k(\As,\Bs)$.

${(4) \Rightarrow (1)}$ Suppose Spoiler plays ${s = [(p_1,a_1),\dots,(p_n,a_n)]}$ in the only round of the game $\exists^{+}\PPeb_k(\As,\Bs)$. We construct a list of Spoiler moves  $(I^{+}_0,I^{-}_1,I^{+}_1,\dots,I^{-}_n,I^{+}_n)$ in $\exists^{+}\PebR_k(\As,\Bs)$, where $I^{+}_{0} = \varnothing$, $I^{-}_i$ is a shrinking move removing the element previously pebbled with $p_i$ from $I^{+}_{i-1}$ and $I^{+}_i$ is a blowing move adding $a_i$ to $I^{-}_i$. Explicitly, 
    \begin{align*}
        I^{-}_i &= I^{+}_{i-1} \backslash \{\last_{p_i}(s[1,i-1])\}; \\
        I^{+}_i &= I^{-}_i \cup \{a_i\}.
    \end{align*}
    By assumption, Duplicator has a winning strategy in $\exists^{+}\PebR_k(\As,\Bs)$, so there exists a list of non-empty partial homomorphisms $(T^{+}_0,T^{-}_1,\dots,T^{-}_n,T^{+}_n)$. Let $\gamma_{1},\dots,\gamma_{n}$ be a list of partial homomorphisms such that $\gamma_{i} \in T^{+}_i$ and $\gamma_{i+1}|_{I^{-}_{i+1}} = \gamma_{i}|_{I^{-}_{i+1}}$.
    Duplicator plays $t = [(p_1,b_1),\dots,(p_n,b_n)]$, where $b_i = \gamma_{i}(a_i)$ in $\exists^{+}\PPeb_k(\As,\Bs)$. This is a winning move for Duplicator since the relation $\gamma_{s[1,i],t[1,i]} = \gamma_i$ by construction.
\end{proof}
We combine Theorem \ref{dalmau2005result} with Theorem \ref{equivgame} to derive the following result:
\begin{cor}\label{existlogiccomonad}
For all $\sigma$-structures $\As$ and $\Bs$,
\[\As \Rrightarrow^{\pRLogicK}\Bs \Leftrightarrow \Rk \As \rightarrow \Bs.\]
\end{cor}
\subsection{Isomorphism power theorem}\label{newlogic}
For the isomorphism power theorem, we first introduce the all-in-one bijective $k$-pebble game $\#\PPeb_{k}(\As,\Bs)$ on $\cala$ and $\calb$, which is a revised bijective analogue of the all-in-one $k$-pebble game.

\begin{defi}[all-in-one bijective $k$-pebble game]
Suppose $\As,\Bs$ are $\sigma$-structures, we define the all-in-one bijective $k$-pebble game $\# \PPeb_k(\As,\Bs)$. During the first and only round,
    \begin{itemize}
    \item Spoiler chooses a sequence of pebbles:
    \[ \vec{p} = [p_1,\dots,p_n]. \]
    \item Duplicator chooses a bijection $h_{\vec{p}}\colon A^{n} \rightarrow B^{n}$.
    \item Spoiler chooses a sequence of pebble placements respecting $\vec{p}$:
    \[ s = [(p_1,a_1),\dots,(p_n,a_n)]  \]
    \item Duplicator must respond with the list of pebble placements:
    \[ t = [(p_1,b_1),\dots,(p_n,b_n)].  \]
    where $h_{\vec{p}}([a_1,\dots,a_n]) = [b_1,\dots,b_n]$.
    \end{itemize}
Duplicator wins the round if for all $i \in \setn$, the relation
    \[\gamma_i = \{ (a^p,b^p) \mid \forall p \in \{p_1,\dots,p_n\}, a^p = \last_{p}(s[1,i]), b^p = \last_{p}(t[1,i])  \} \]
is a partial isomorphism.
\end{defi}

As previously mentioned, we work with the bijective variant $\#\PPeb_k(\As,\Bs)$ of the all-in-one $k$-pebble game $\exists^{+}\PPeb_k(\As,\Bs)$, since it is much easier to work with and generalise, in our context, than the $k$-pebble relation game.

The proof of this theorem follows similar lines to the analogous proof for the bijective Ehrenfeucht-{\Fraisse} game given by \mbox{Libkin \cite{libkin}}.

We define the logic $\cRLogicK$ where the equivalence relation $\equiv_{\cRLogicK}$ is characterised by Duplicator winning strategies in $\#\PPeb_k(\As,\Bs)$ and thus, isomorphism in $\Kl(\Rk)$. 

In order to define our logic $\cRLogicK$, we first define a logic $\eRLogicK(Z)$. 
This logic is the same as $\pRLogicK$, but in addition to the $k$-many ordinary variables $X = \{x_1,\dots,x_k\}$, we have an another sort of infinitely-many variables $Z = \{z_1,z_2,\dots\}$ which we call `walk pointer variables'. 
We also allow negations on atomic formulas in $\eRLogicK(Z)$ and require that every existential quantifier $\exists x_j$ is guarded with an equality $x_j = z_i$ equating a ordinary variable $x_j$ with a walk pointer variable $z_i$.
Explicitly, the formulas of $\eRLogicK(Z)$ can be defined recursively as follows:
\[\psi(\vec{x},\vec{z}) ::=R(x_{i_1},\dots,x_{i_r})\;|\;\neg p\;|\;\bigwedge\Psi\;|\;\bigvee\Psi\;|\;\exists x_j(x_j = z_i \wedge \psi(\vec{x'},\vec{z}')),\]
where $R \in \sigma$ is an $r$-ary relation such that $\{x_{i_1},\dots,x_{i_r}\} \in [\vec{x}] \subseteq \{x_1,\dots,x_k\}$, $p$ is an atomic formula, $\bigwedge\Psi$ is a conjunction of formulas $\psi(\vec{x},\vec{z})$ satisfying~\ref{linearCond}, $\bigvee\Psi$ is a disjunction of formulas $\psi(\vec{x},\vec{z})$, $x_j\in[\vec{x'}]$ and $z_i\notin \vec{z}'$.

We can now define sentences of $\cRLogicK$ as disjunctions of formulas that involve counting tuples of $z$-variables. 
Explicitly, the sentence of $\cRLogicK$ can be defined recursively as follows:
\[  \phi ::=\bigvee \Phi\;|\;\exists^j (z_{i_1},\dots,z_{i_m})\psi(z_{i_1},\dots,z_{i_m}),\]
where $\Phi$ is a collection of sentences $\phi$, $j \in \mathbb{N}$ and $\psi(z_{i_1},\dots,z_{i_m}) \in \eRLogicK(Z)$.

Observe that in the syntax of the $\exists^j \vec{z}$ quantifier every free variable in the bound formula $\psi(\vec{z}) \in \eRLogicK(Z)$ is from the sort $Z$ of walk pointer variables.
We define the semantics of the operation above as
\begin{itemize}
    \item      $\As \vDash \exists^j (z_{i_1},\dots,z_{i_m})\psi(z_{i_1},\dots,z_{i_m})$ iff there exist exactly $j$-many tuples $\vec{a} = (a_1,\dots,a_m)$ such that $\As,\vec{a} \vDash \psi(z_{i_1},\dots,z_{i_m})$.
\end{itemize}
The quantifier $\exists^j (z_{i_1},\dots,z_{i_m})$ in the logic $\cRLogicK$ is inspired by a similar $k$-walk quantifier used in the logic defined in~\cite{walkCounting}.

A formula of $\varphi(\vec{x}) \in \pRLogicK$ is primitive if it contains no disjunctive subformula and every restricted conjunction subformula $\bigvee \Psi$ does not contain a sentence. 
From definition of primitive formula and the condition \ref{linearCond}, it follows that every conjunctive subformula $\bigvee \Psi$ of a primitive formula $\varphi(x)$ contains only literals and at most one quantified formula with a free variable.
We similarly can define primitive formulas of $\eRLogicK(Z)$. 
Thus, we extend this definition to $\cRLogicK$, and say a sentence of $\phi \in \cRLogicK$ is \emph{primitive} if it is of the form $\phi = \exists^j (z_{i_1},\dots,z_{i_m})\psi(z_{i_1},\dots,z_{i_m})$ if $\psi(\vec{z}) \in \eRLogicK(Z)$ is primitive. 
\begin{prop}
\label{prop:primitive-form}
Every sentence $\phi \in \cRLogicK$ is a disjunction of primitive subformulas.
\end{prop}

\begin{proof}
Given a sentence $\phi$, we can inductively apply the standard rewrite rule 
\[\exists y(\bigvee_{i \in I} \varphi(y,\vec{x})) \mapsto\bigvee_{i \in I} \exists y \varphi(y,\vec{x})\] 
along with commutativity of $\bigwedge,\bigvee$ to obtain an equivalent formula. 
From the sematics it is clear that this standard rewrite rule also works for the walk counting quantifers $\exists^{j} (z_{i_1},\dots,z_{i_m})$.
\end{proof}
We now prove the equivalence between truth preservation in $\cRLogicK$ and the game $\#\PPeb_k(\As,\Bs)$.

\begin{thm}\label{gamelogic}
The following are equivalent for every $\sigma$-structures $\As$ and $\Bs$:
\begin{enumerate}
	\item Duplicator has a winning strategy in $\#\PPeb_k(\As,\Bs)$.
	\item $\As \equiv^{\cRLogicK}\Bs$.
	\end{enumerate}	
\end{thm}
\begin{proof}
$\Rightarrow$ Suppose Duplicator has a winning strategy in $\#\PPeb_k(\As,\Bs)$. Consider a primitive sentence $\phi$ of $\cRLogicK$ such that $\As \vDash \phi$. As $\phi$ is primitive, it is of the form $\exists^{j}(z_{i_1},\dots,z_{i_m})\psi(z_{i_1},\dots,z_{i_m})$ for some $j \in \mathbb{N}$ and primitive $\psi(z_{i_1},\dots,z_{i_m}) \in \eRLogicK(Z)$.
From $\psi$, we construct Spoiler's sequence of pebbles $\vec{p} = [p_1,\dots,p_n]$ in the game $\#\PPeb_k(\As,\Bs)$ such that for all $l \in \setm$, $x_{p_l}  = z_{i_l}$ is subformula of $\psi$ guarding a existential quantifier.
Let $h_{\vec{p}}\colon A^{m} \rightarrow B^{m}$ denote Duplicator response in her winning strategy for $\#\PPeb_k(\As,\Bs)$.
By $\As \vDash \phi$, there exist exactly $j$-many tuples $\vec{a} = (a_1,\dots,a_m)$ such that $\As,\vec{a} \vDash \psi(z_{i_1},\dots,z_{i_m})$.
Thus, for each of these tuples, we examine the play where Spoiler plays the sequence $[(p_1,a_1),\dots,(p_m,a_m)]$.
Duplicator responds with the sequence $[(p_1,b_1),\dots,(p_m,b_m)]$ such that $h_{\vec{p}}(\vec{a}) = \vec{b} = [b_1,\dots,b_m]$.
From the winning condition it follows that $\Bs,\vec{b} \vDash \psi(z_{i_1},\dots,z_{i_m})$.
Since $h_{\vec{p}}$ is a bijection there are $j$-many tuples $\vec{b}$ such that $\Bs,\vec{b} \vDash \psi(z_{i_1},\dots,z_{i_m})$.
Thus, $\Bs \vDash \phi$.
A similar argument shows that if $\Bs \vDash \phi$, then $\As \vDash \phi$.
By Proposition~\ref{prop:primitive-form}, every sentence in $\cRLogicK$ is a disjunction of primitive sentences $\phi$.
Therefore, $\As \equiv_{\cRLogicK} \Bs$.

$\Leftarrow$ Before we proceed with this proof, we will need to define some formulas that are associated with plays in $\#\PPeb_k(\As,\Bs)$. 

First, we define, for every sequence $s = [(p_1,a_1),\dots,(p_n,a_n)]$ of pebble placements on $A$, the literal diagram of $s$ as 
\begin{align*}
  \diag_{\As}(s) &= \{ R(\mathbf{x}(p_{i_1}),\dots,\mathbf{x}(p_{i_r}))\mid \forall R \in \sigma, \forall z \in \setr, \last_{p_{i_z}}(s) = a_{i_z} \wedge (a_{i_1},\dots,a_{i_r}) \in R^{\As} \} \\
           &\cup \{ \neg R(\mathbf{x}(p_{i_1}),\dots,\mathbf{x}(p_{i_r})) \mid \forall R \in \sigma, \forall z \in \setr, \last_{p_{i_z}}(s) = a_{i_z} \wedge (a_{i_1},\dots,a_{i_r}) \not\in R^{\As}\}
\end{align*}
where $\mathbf{x}(p) = x_{p}$, i.e.\ $\mathbf{x}$ returns the ordinary $x$ variable corresponding to the pebble $p \in \setk$.
Next, for a sequence of pebble placements $s = [(p_1,a_1),\dots,(p_n,a_n)]$ of length $n$, we define the formula $\tp_{s}(z_1,\dots,z_n) \in \eRLogicK(Z)$ with walk pointer free variables $\{z_1,\dots,z_n\} \subseteq Z$ such that $\As,(a_1 \dots a_n) \vDash \tp_{s}(z_1,\dots,z_n)$.
To define $\tp_{s}$, for a sequence of pebble placements $s = [(p_1,a_1),\dots,(p_n,a_n)]$ of length $n$, let $m \in \{0,\dots,n-1\}$ be the largest index such that $p_m = p_n$.
By induction on $j$ up to $m$, we define the formulas $\varphi_{j}(\vec{z},\vec{x})$, where $\vec{z} = (z_1,\dots,z_j)$ are walk-pointer variables, and $\vec{x} = (\mathbf{x}(p_{1}),\dots,\mathbf{x}(p_{j-1}))$ are ordinary variables corresponding to the pebbles $\{p_{1},\dots,p_{j-1}\}$ in $\pebbles(s[1,j]) \backslash \{p_j\}$:
\begin{align*}
\varphi_{0} &= \top; \\ 
\varphi_{j}(z_{j}\vec{z},\vec{x}) &= \exists \mathbf{x}(p_j)(z_{j} = \mathbf{x}(p_j) \wedge \bigwedge \diag_{\As}(s[1,j]) \wedge \varphi_{j-1}(\vec{z},\vec{x'}));
\end{align*}
where $\mathbf{x}(p_j)$ is possibly in $\vec{x}'$.
Finally, we define 
\[ \tp_{s}(z_1,\dots,z_n) := \exists \vec{x}\left( \bigwedge_{j \in \{m+1,\dots,n\}} z_{j} = \mathbf{x}(p_j) \wedge \bigwedge \diag_{\As}(s) \wedge \varphi_{m}(\vec{z'},\vec{x'})\right), \]  
where $\vec{z'} = (z_1,\dots,z_m)$, $\vec{x} = (\mathbf{x}(p_{m+1}),\dots,\mathbf{x}(p_n))$ and $\exists \vec{x}:=\exists \mathbf{x}(p_{m+1}) \dots \exists \mathbf{x}(p_n)$.

We now proceed with the proof using these type formulas $\tp_s$. 
Suppose by hypothesis $\As \equiv_{\cRLogicK} \Bs$. 
In particular, for every sentence of the form $\phi = \exists^{j}(z_1,\dots,z_n)\tp_{s}(z_1,\dots,z_n)$, $\As \vDash \phi \Leftrightarrow \Bs \vDash \phi$.
This determines a bijection $h\colon A^{n} \rightarrow B^{n}$ such that:
\[ \As,(a_1,\dots,a_n) \vDash \tp_{s}(z_1,\dots,z_n) \Leftrightarrow \Bs,(h(b_1),\dots,h(b_n)) \vDash \tp_{s}(z_1,\dots,z_n) \]
By construction of $\tp_{s}(z_1,\dots,z_n)$, this yields a Duplicator winning strategy in\linebreak $\#\PPeb_k(\As,\Bs)$. 
Namely, for all indices $i \in \setn$, either $i \leq m$ and $\diag_{\As}(s[1,i])$ is a subformula of $\tp_{s}(z_1,\dots,z_n)$, or $i > m$ and $\diag_{\As}(s[1,i]) \subseteq \diag_{\As}(s)$.
Therefore, $\As,(a_j,\dots,a_i) \vDash \bigwedge \diag_{\As}(s[1,i])$ if and only if $\Bs,(h(a_j),\dots,h(a_n)) \vDash \bigwedge \diag_{\As}(s[1,i])$, where $\Active(s[1,i]) = \{a_j,\dots,a_i\}$.
Thus, since $\bigwedge \diag_{\As}(s[1,i])$ expresses the satisfaction of all literals in \linebreak$\{a_j,\dots,a_i\} \subseteq A$ and $\{h(a_j),\dots,h(a_i)\} \subseteq B$, $\gamma_{i}$ is a partial isomorphism from $\As$ to $\Bs$.
\end{proof}

We prove the correspondence Kleisli isomorphism in $\Kl(\Rk)$ and $\#\PPeb_k(\As,\Bs)$.
Given a pebble sequence of some finite length $[p_1,\dots,p_n] \in \setk^{\leq \omega}$ and $\sigma$-structure $\As$, let $P_{\vec{p}}(\As)$ be the induced substructure of $\Rk(\As)$ on the set 
\[ P_{\vec{p}}(A) = \{([(p_1,a_1),\dots,(p_n,a_n)],i) \mid a_1,\dots,a_n \in A \text{ and } i \in \setn\} \]
Since the coextension $f^{*}\colon \Rk \As \rightarrow \Rk \Bs$ of a Kleisli morphism $f^{*}\colon \Rk \As \rightarrow \Bs$ sends sequences of pebble placements on $\As$ paired with an index to compatible lists of pebble placements on $\Bs$ paired with the same index, for every $\vec{p} \in \setk^{\leq \omega}$, the restriction of $f^{*}$ induces a $\sigma$-morphism $f_{\vec{p}}\colon P_{\vec{p}}(\As) \rightarrow P_{\vec{p}}(\Bs)$.
We also note that if $f\colon \Rk \As \rightarrow \Bs$ is an isomorphism, then $f_{\vec{p}} \colon P_{\vec{p}}(\As) \rightarrow P_{\vec{p}}(\Bs)$ is an isomorphism.
\begin{thm}
\label{bilogcom}
The following are equivalent:
\begin{enumerate}
    \item There exists a Kleisli isomorphism $\Rk J \As \rightarrow J \Bs$ 
    \item Duplicator has a winning strategy in $\# \PPeb_k(\As,\Bs)$
\end{enumerate}
\end{thm}
\begin{proof}
$\Rightarrow$ Assume there exists a Kleisli isomorphism $f\colon \Rk J \As \rightarrow \Bs$. We need construct a winning strategy for Duplicator in $\# \PPeb_k(\As,\Bs)$, so suppose Spoiler plays the sequence $\vec{p} = [p_1,\dots,p_n]$.
The coextension $f^{*}$ of the Kleisli isomorphism $f$ when restricted to $P_{\vec{p}}(\As)$ is the isomorphism $f_{\vec{p}} \colon P_{\vec{p}}(\As) \rightarrow P_{\vec{p}}(\Bs)$. 
For every $[a_1,\dots,a_n]$, there exists an $[b_1,\dots,b_n]$, such that for all $i \in \setn$, $f_{\vec{p}}(([(p_1,a_1),\dots,(p_n,a_n)],i)) = ([(p_1,b_1),\dots,(p_n,b_n)],i)$. 
We define Duplicator's response to be the function $\psi\colon A^{n} \rightarrow B^n$ which maps $[a_1,\dots,a_n]$ to $[b_1,\dots,b_n]$. 

It follows from $f_{\vec{p}}$ being an isomorphism that $\psi$ is a bijection.
Namely. consider $g_{\vec{p}}\colon P_{\vec{p}}(\Bs) \rightarrow P_{\vec{p}}(\As)$ to generated from the inverse $g^{*}\colon \Rk J \Bs \rightarrow \Rk J \As$ of $f^{*}$.
There exists a $\phi\colon B^n \rightarrow A^n$ such that $\phi[b_1,\dots,b_n] = [a_1,\dots,a_n]$ if and only for all $i \in \setn$ $g_{\vec{p}}([(p_1,b_1),\dots,(p_n,b_n)],i) = ([(p_1,a_1),\dots,(p_n,a_n)],i)$.
Clearly, by construction $\psi^{-1} = \phi$.

Moreover, if Spoiler plays the sequence $s = [(p_1,a_1),\dots,(p_n,a_n)]$, then Duplicator must respond with $t = [(p_1,b_1),\dots,(p_n,b_n)]$ as $\psi[a_1,\dots,a_n] = [b_1,\dots,b_n]$.
These sequences determine relation $\gamma_i$ for all $i \in \setn$.
By $f_{\vec{p}}$ being an isomorphism, we can show that
$\gamma_i$ is a partial isomorphism.
Suppose $R \in \sigma$ and consider the pairs $(a^{q_1},b^{q_1}),\dots,(a^{q_r},b^{q_r}) \in \gamma_i$ such that $R^{\As}(a^{q_1},\dots,a^{q_r})$.
By definition of $\gamma_i$, each $q_{j} = \pi_{\As}(s,z_j)$ and $a^{q_j} = \varepsilon_{\As}(s,j)$ for some $z_j \leq i \leq n$. 
Thus, by $R^{\As}(a^{q_1},\dots,a^{q_r})$, we have that $R^{\As}(\varepsilon(s,z_1),\dots,\varepsilon(s,z_r))$ and $R^{\Rk(\As)}((s,z_1),\dots,(s,z_r))$. 
By $f_{\vec{p}}$ being a morphism, $R^{\Rk(\Bs)}(f_{\vec{p}}(s,z_1),\dots,f_{\vec{p}}(s,z_r))$ and $R^{\Rk(\Bs)}((t,z_1),\dots,(t,z_r))$. 
Applying $\varepsilon_{\Bs}$, we obtain that $R^{\Rk(\Bs)}((t,z_1),\dots,(t,z_r))$ and $R^{\Bs}(b^{q_1},\dots,b^{q_r})$.
A similar proof, using the inverse of $f_{\vec{p}}$, show that if  $R^{\Bs}(b^{q_1},\dots,b^{q_r})$, then $R^{\As}(a^{q_1},\dots,a^{q_r})$.
Thus, $\gamma_i$ is partial correspondence and using the fact that $J\As$ and $J\Bs$ interpret $I$ as equality, we obtain that $\gamma_i$ is a partial isomorphism for all $i \in \setn$.

$\Leftarrow$ Conversely, suppose Duplicator has a winning strategy in $\#\PPeb_k$. 
Suppose Spoiler plays the pebble sequence $\vec{p} = [p_1,\dots,p_n]$ and Duplicator responds with $\psi_{\vec{p}}\colon A^n \rightarrow B^n$. 
We define the $\sigma$-isomorphism $f_{\vec{p}}\colon P_{\vec{p}}(\As) \rightarrow P_{\vec{p}}(\Bs)$ such that if $\psi_{\vec{p}}[a_1,\dots,a_n] = [b_1,\dots,b_n]$, then $f_{\vec{p}}([(p_1,a_1),\dots,(p_n,a_n)],i) = ([(p_1,b_1),\dots,(p_n,b_n)],i)$ for all $i \in \setn$.
Taking the coproduct of $f_{\vec{p}}\colon P_{\vec{p}}(\As) \rightarrow P_{\vec{p}}(\Bs)$ over all $\vec{p} \in \setk^{< \omega}$ we obtain the isomorphism $f^{*}\colon \Rk J \As \rightarrow \Rk J \Bs$. 
Composing with $\varepsilon_{J \Bs}$ yields the existence of the Kleisli isomorphism $f\colon \Rk J \As \rightarrow \Bs$.
\end{proof}

We combine the above theorems to derive the following result:
\begin{cor}\label{logiccomonad}
For all finite $\sigma$-structures $\As$ and $\Bs$,
\[\As \equiv^{\cRLogicK}\Bs \Leftrightarrow J\As \cong^{\Kl(\Rk)} J\Bs.\] 
\end{cor}
If we restrict the proofs of Theorems \ref{gamelogic} and \ref{bilogcom} to only sequences $s$ of size up to $n$, we obtain a revised statement of Corollary \ref{logiccomonad} expressing that equivalence in the quantifier rank $\leq n$ fragment of $\cRLogicK$ is characterised by a coKleisli isomorphism of $\Rkn$.

One of the contributions of the pebbling comonad was unifying two algorithms: the $k$-Weisfeiler-Lehman algorithm deciding the existence of a coKleisli isomorphism $J\As \cong^{\Kl(\Pk)} J\Bs$ and the $k$-consistency algorithm deciding the existence of a coKleisli morphism $\Pk \As \rightarrow \Bs$, where $\As$ and $\Bs$ represent the variables and values (respectively) of a given CSP \cite{abramsky2017}.   
Corollary \ref{logiccomonad} suggests that understanding algorithms for deciding $\Rk \As \rightarrow \Bs$ may provide insights into how to construct algorithms for deciding $J\As \cong^{\Kl(\Rk)} J\Bs$ (and vice-versa).
We briefly discuss the complexity of such algorithms in the conclusion.

\subsection{{\Lovasz}-type theorem for pathwidth}
In one of his seminal papers, {\Lovasz} \cite{lovsz1967} proved that two finite $\sigma$-structures $\As$ and $\Bs$ are isomorphic if and only if for all finite $\sigma$-structures $\Cs$, the number of homomorphisms from $\Cs$ to $\As$ is the same as the number of homomorphisms from $\Cs$ to $\Bs$. 
Following this result, weaker notions of equivalence have been found to have an analogous characterisation \cite{dvok2009,grohe2020,manvcinska2020}. 
These equivalences involve restricting the class of test structures $\Cs$.
For instance, Dvo{\v{r}}\'ak  \cite{dvok2009} showed that counting homomorphisms from finite $\sigma$-structures of treewidth $< k$ yields equivalence in $k$-variable logic with counting quantifiers. 

 Counting homomorphisms from graphs of treewidth $< k$ also has a linear algebraic characterisation in terms of existence of a non-negative real solution to a certain system of linear \mbox{equations \cite{holger2018}.} 
It was shown in a later work by Grohe, Rattan and Seppelt \cite{grohe2021} that removing this non-negativity constraint yields a homomorphism counting result from graphs of pathwidth $< k$. 
In their work they used techniques from representation theory.
By contrast, we use techniques from category theory to obtain an analogue of Dvo{\v{r}}\'ak's result providing a logical characterisation of homomorphism counting from $\sigma$-structures with pathwidth $< k$.

In recent work, Dawar, Jakl and Reggio \cite{dawar2021} introduced a generalisation of {\Lovasz}-type theorems for comonads over $\Rsig$. 
In this section we utilise this generalisation in order to show that counting homomorphisms from finite $\sigma$-structures of pathwidth $< k$ yields equivalence in restricted conjunction $k$-variable logic with counting quantifiers. 

We begin by introducing the categorical generalisation for {\Lovasz}-type theorems \cite{dawar2021}. 

Let $\CC_k$ and $\CC^+_k$ be comonads on $\Rsig$ and $\Rsigplus$, respectively. 
These comonads yield the forgetful-cofree adjunctions $U^{\CC_k} \dashv F^{\CC_k}$ and $U^{\CC_k^+} \dashv F^{\CC_k^+}$ where $U^{\CC_k}:\EM(\CC_k) \rightarrow \Rsig$ and $F^{\CC_k}:\Rsigplus \rightarrow \EM(\CC_k)$ (similarly for $\CC_k^+$).
Let $\CC^+_k J$ be the relative comonad on $J\colon \Rsig \rightarrow \Rsigplus$.
The functor $J$ has a left adjoint $H:\Rsigplus \rightarrow \Rsig$ sending $\As \in \Rsigplus$ to the quotient structure ${\As^{-}/\sim} \in \Rsig$, where $\As^{-}$ is the $\sigma$-reduct of $\As$ and $\sim$ is the least equivalence relation generated from $I^{\As}$.
Below, recall that $\Rsigf$ denotes the full subcategory of $\sigma$-structures with finite universes.
\begin{thmC}[\cite{dawar2021}]\label{lovasz}
	Given a logic $\Logic_k$, suppose that the following three conditions are satisfied:
	\begin{enumerate}
		\item For all finite $\sigma$-structures $\cala$ and $\calb$ 
		\[ \cala\equiv_{\Logic_k} \calb \iff  F^{\CC_k^{+}}(\As)\cong F^{\CC_k^{+}}(\Bs);\]
		\item $\CC^+_k J$ sends finite structures in $\Rsig$ to finite structures in $\Rsigplus$; 
		\item $J:\Rsig\rightarrow\Rsigplus$ and $H:\Rsigplus\rightarrow\Rsig$ restrict to the subcategories of finite relational structures $\Rsigf\cap \emph{im}(U^{\CC_k})$ and $\Rsigplusf\cap \emph{im}(U^{\CC^+_k})$. 
	\end{enumerate}
	Then for every finite $\sigma$-structures $\cala$ and $\calb$,
	\[\cala\equiv_{\Logic_k}\calb \iff |{\Rsigf}(\mathcal C,\cala)|=|{\Rsigf}(\mathcal C,\calb)|,\]
	for all finite $\sigma$-structure $\mathcal C$ in $\emph{im}(U^{\CC_k})$.
	\label{thm:comonadLovasz}
\end{thmC}
The categorical interpretation of pathwidth and the isomorphism power theorem for $\Rk$ allow us to derive the following consequence from Theorem \ref{lovasz}:
\begin{cor}[{\Lovasz}-type theorem for pathwidth]\label{lovasz2}
For all finite $\sigma$-structures $\cala$ and $\calb$,
\[ \cala\equiv^{\cRLogicK}\calb\iff |{\Rsigf}(\mathcal C,\cala)|=|{\Rsigf}(\mathcal C,\calb)|,\]
for every finite $\sigma$-structure $\mathcal C$ with pathwidth $< k$.
\end{cor}
Intuitively, since $\Rk\As$ is infinite for every structure $\As$, in order to meet condition 2 we employ a similar technique to the one Dawar et al.\ used when applying Theorem \ref{thm:comonadLovasz} to $\Pk$. We consider the pebble-relation comonad $\Rkn$ that is also graded by sequence length. 
Recall that $\Rkn\As$ consists of pairs $(s,i)$, where $s$ is a sequence of length at most $n$. 
Each of the three conditions follows from the results in \cite{dawar2021} applied to $\Rkn$.

Condition 1 follows from Corollary \ref{logiccomonad} when applied to the comonad $\Rkn$, where $F^{\CC_k^{+}}$ is instantiated as the cofree functor $F_{k,n}:\Rsigplus \rightarrow \EM(\Rkn)$ from the adjunction $U_{k,n} \dashv F_{k,n}$.

Condition 2 follows from the fact that $\Rkn \As$ is finite whenever $\As$ is finite.

Condition 3 requires that if $\As$ has a $k$-pebble linear forest cover, then $J\As$ and $HJ\As$ have $k$-pebble linear forest covers. 
The paper by Dawar et al.\ proves this statement for general $k$-pebble forest covers (Lemma 22 of \cite{dawar2021}) as part of their application of Theorem \ref{thm:comonadLovasz} to $\Pk$. 
In particular, their argument holds for paths in the category of $k$-pebble forest covers.
Therefore, the statement holds for coproducts of paths, i.e.\ $k$-pebble linear forest covers.

Finally, realising $\Rk$ as the colimit of $\Rkn$ for all $n \in \omega$ allows us to obtain the desired result. 

\section{Conclusion}
\label{sec:conclusion}
The results we exhibited extend the growing list of Spoiler-Duplicator game comonads that unify particular model-comparison games with combinatorial invariants of relational structures. 
This approach has applications in reformulations of Rossman's homomorphism preservation theorem \cite{paine2020}, and it provides a new perspective on finite model theory and descriptive complexity. 
In particular, $\Rk$ provides a categorical definition for pathwidth as well as winning strategies in Dalmau's pebble-relation game and the all-in-one pebble game.
This allowed us to obtain a syntax-free characterisation of equivalence in the restricted conjunction fragment of $k$-variable logic. 

The comonad $\Rk$ was obtained in a unique way by first trying to capture the combinatorial invariant, instead of internalising the corresponding game, as was done for previously studied game comonads \cite{abramsky2017,abramskyResources2018,oconghaile2020,abramsky2020}.
Moreover, some of the proofs in this paper use techniques distinct from those used for other Spoiler-Duplicator comonads (e.g.\ $k$-pebbling section families, fibres over $\Pk\As$ to model non-determinism). 
The coKleisli isomorphism of $\Rk$ allowed us to define an original bijective game that characterises equivalence in the heretofore unexplored restricted conjunction fragment of $k$-variable logic with counting quantifiers. 
Finally, we proved a {\Lovasz}-type theorem linking equivalence in this logic with homomorphism counting from structures of pathwidth $< k$. 

We share a few possible avenues for future work:
\begin{itemize}
    \item Dalmau \cite{dalmau2005} claims that constraint satisfaction problems (CSPs) with bounded pathwidth duality are in $\textbf{NL}$. All known CSPs in $\textbf{NL}$ have bounded pathwidth duality. 
    One possible inquiry would be to check if redefining bounded pathwidth duality in terms of $\Rk$ can help in finding a proof for the converse or a construction of a counterexample.
    
    Egri \cite{egri2014} proved that 
    CSPs that have bounded symmetric pathwidth duality are in $\textbf{L}$. 
    Whereas bounded pathwidth duality can be seen as a local property of the obstruction set, bounded symmetric pathwidth duality is a global property of the obstruction set. 
    By reformulating bounded symmetric pathwidth duality in terms of $\Rk$, we can further reinforce the comonadic analysis to understand the complexity of CSPs.  
    
    \item For the pebbling comonad $\Pk$, we can interpret a coKleisli morphism $\Pk\As \rightarrow \Bs$ as a simulation between Kripke models constructed from the sets of tuples $A^{\leq k}$ and $B^{\leq k}$.  
    Analogously, a coKleisli morphism $\Rk \As \rightarrow \Bs$ can be interpreted as trace inclusion on these very same Kripke models. 
    Intuitively, Duplicator's response in the all-in-one $k$-pebble game maps a trace of the Kripke model built from  $A^{\leq k}$ to a trace of the Kripke model built from $B^{\leq k}$.
    While simulation of Kripke models is $P$-complete, trace inclusion\footnote{This is originally called trace preorder in \cite{huttel1996}.} is \mbox{$PSPACE$-complete \cite{huttel1996},} leading to the following conjecture:
    \begin{conj}
    Deciding $\Rk \As \rightarrow \Bs$ for inputs $\As,\Bs$ and fixed $k$ is $PSPACE$-complete. 
    \end{conj}
    
    \item The pebble-relation comonad $\Rk$ can be seen as the `linear', or path variant, of the `tree shaped' pebbling comonad $\Pk$. 
    In fact, as the notion of arboreal categories discussed in \cite{abramsky2021} shows, all of the previously discovered Spoiler-Duplicator game comonads $\CC_k$ are `tree shaped', i.e.\ $\EM(\CC_k)$ is arboreal.
    One may investigate whether there exists a general method for obtaining a `linear' variant of a `tree shaped' Spoiler-Duplicator game comonad $\CC_k$.
    Moreover, the framework of arboreal categories demonstrates that a certain categorical generalisation of bisimulation in $\EM(\CC_k)$ captures equivalence in $\Logic_k$ associated with $\CC_k$. 
    Discovering which fragment of $\LogicK$ is captured by a similar relation in $\EM(\Rk)$ remains an open question.
\end{itemize}

\bibliographystyle{alphaurl}
\bibliography{main}

\begin{thebibliography}{ADW17}

\bibitem[ACU14]{altenkirch2014}
Thosten Altenkirch, James Chapman, and Tarmo Uustalu.
\newblock Monads need not be endofunctors.
\newblock {\em Logical Methods in Computer Science}, 11(1), 2014.

\bibitem[ADW17]{abramsky2017}
Samson Abramsky, Anuj Dawar, and Pengming Wang.
\newblock The pebbling comonad in finite model theory.
\newblock In {\em Logic in Computer Science (LICS), 2017 32nd Annual ACM/IEEE
  Symposium on}, pages 1--12. IEEE, 2017.

\bibitem[AM21]{abramsky2020}
Samson Abramsky and Dan Marsden.
\newblock Comonadic semantics for guarded fragments.
\newblock In {\em 2021 36th Annual {ACM}/{IEEE} Symposium on Logic in Computer
  Science ({LICS})}. {IEEE}, June 2021.

\bibitem[AR21]{abramsky2021}
Samson Abramsky and Luca Reggio.
\newblock {Arboreal Categories and Resources}.
\newblock In Nikhil Bansal, Emanuela Merelli, and James Worrell, editors, {\em
  48th International Colloquium on Automata, Languages, and Programming (ICALP
  2021)}, volume 198 of {\em Leibniz International Proceedings in Informatics
  (LIPIcs)}, pages 115:1--115:20, Dagstuhl, Germany, 2021. Schloss Dagstuhl --
  Leibniz-Zentrum f{\"u}r Informatik.

\bibitem[AS18]{abramskyResources2018}
Samson Abramsky and Nihil Shah.
\newblock Relating structure and power: Comonadic semantics for computational
  resources.
\newblock In {\em 27th {EACSL} Annual Conference on Computer Science Logic,
  {CSL} 2018, September 4-7, 2018, Birmingham, {UK}}, pages 2:1--2:17, 2018.

\bibitem[AS21]{abramskyResources2021}
Samson Abramsky and Nihil Shah.
\newblock Relating structure and power: Comonadic semantics for computational
  resources.
\newblock {\em Journal of Logic and Computation}, 31(6):1390--1428, August
  2021.

\bibitem[CD21]{oconghaile2020}
Adam~{\'O} Conghaile and Anuj Dawar.
\newblock Game comonads \& generalised quantifiers.
\newblock In {\em 29th EACSL Annual Conference on Computer Science Logic (CSL
  2021)}. Schloss Dagstuhl-Leibniz-Zentrum f{\"u}r Informatik, 2021.

\bibitem[Dal05]{dalmau2005}
Victor Dalmau.
\newblock Linear {Datalog} and {Bounded} {Path} {Duality} of {Relational}
  {Structures}.
\newblock {\em Logical Methods in Computer Science}, 1(1):5, April 2005.

\bibitem[DGR18]{holger2018}
Holger Dell, Martin Grohe, and Gaurav Rattan.
\newblock Lov{\'{a}}sz meets weisfeiler and leman.
\newblock In Ioannis Chatzigiannakis, Christos Kaklamanis, D{\'{a}}niel Marx,
  and Donald Sannella, editors, {\em 45th International Colloquium on Automata,
  Languages, and Programming, {ICALP} 2018, July 9-13, 2018, Prague, Czech
  Republic}, volume 107 of {\em LIPIcs}, pages 40:1--40:14. Schloss Dagstuhl -
  Leibniz-Zentrum f{\"{u}}r Informatik, 2018.

\bibitem[DJR21]{dawar2021}
Anuj Dawar, Tomas Jakl, and Luca Reggio.
\newblock Lov{\'{a}}sz-type theorems and game comonads.
\newblock In {\em 2021 36th Annual {ACM}/{IEEE} Symposium on Logic in Computer
  Science ({LICS})}. {IEEE}, June 2021.

\bibitem[Dvo09]{dvok2009}
Zden{\v{e}}k Dvo{\v{r}}{\'{a}}k.
\newblock On recognizing graphs by numbers of homomorphisms.
\newblock {\em Journal of Graph Theory}, 64(4):330--342, November 2009.

\bibitem[Egr14]{egri2014}
L{\'a}szl{\'o} Egri.
\newblock On constraint satisfaction problems below p.
\newblock {\em Journal of Logic and Computation}, 26(3):893--922, 2014.

\bibitem[Gro17]{grohe2017}
Martin Grohe.
\newblock {\em Descriptive Complexity, Canonisation, and Definable Graph
  Structure Theory}.
\newblock Cambridge University Press, 2017.

\bibitem[Gro20]{grohe2020}
Martin Grohe.
\newblock Counting bounded tree depth homomorphisms.
\newblock In {\em Proceedings of the 35th Annual {ACM}/{IEEE} Symposium on
  Logic in Computer Science (LICS)}. {ACM}, July 2020.

\bibitem[GRS22]{grohe2021}
Martin Grohe, Gaurav Rattan, and Tim Seppelt.
\newblock Homomorphism tensors and linear equations, 2022.
\newblock To appear in \emph{49th EATCS International Colloquium on Automata,
  Languages and Programming (ICALP 2022)}.

\bibitem[Hel96]{hella1996}
Lauri Hella.
\newblock {Logical hierarchies in PTIME}.
\newblock {\em Information and Computation}, 121:1--19, 1996.

\bibitem[HS96]{huttel1996}
Hans H\"uttel and Sandeep Shukla.
\newblock On the complexity of deciding behavioural equivalences and preorders.
\newblock Technical report, USA, 1996.

\bibitem[Imm82]{immerman1982}
Neil Immerman.
\newblock {Upper and lower bounds for first order expressibility}.
\newblock {\em Journal of Computer and System Sciences}, 28:76--98, 1982.

\bibitem[KV90]{kolaitis1990}
Phokion~G Kolaitis and Moshe~Y Vardi.
\newblock {On the expressive power of Datalog: tools and a case study}.
\newblock In {\em Proceedings of the ninth ACM SIGACT-SIGMOD-SIGART symposium
  on Principles of database systems}, pages 61--71. ACM, 1990.

\bibitem[Lib04]{libkin}
Leonid Libkin.
\newblock {\em {Elements of Finite Model Theory (Texts in Theoretical Computer
  Science. An EATCS Series)}}.
\newblock Springer, 2004.

\bibitem[Lov67]{lovsz1967}
L{\'a}szl{\'o} Lov{\'{a}}sz.
\newblock Operations with structures.
\newblock {\em Acta Mathematica Academiae Scientiarum Hungaricae},
  18(3-4):321--328, September 1967.

\bibitem[LPS19]{walkCounting}
Moritz Lichter, Ilia Ponomarenko, and Pascal Schweitzer.
\newblock Walk refinement, walk logic, and the iteration number of the
  weisfeiler-leman algorithm.
\newblock In {\em 2019 34th Annual ACM/IEEE Symposium on Logic in Computer
  Science (LICS)}, pages 1--13, 2019.
\newblock \href {https://doi.org/10.1109/LICS.2019.8785694}
  {\path{doi:10.1109/LICS.2019.8785694}}.

\bibitem[Man76]{manes1976}
Ernest~G. Manes.
\newblock {\em Algebraic Theories}.
\newblock Springer New York, 1976.

\bibitem[MR20]{manvcinska2020}
Laura Man{\v{c}}inska and David~E Roberson.
\newblock Quantum isomorphism is equivalent to equality of homomorphism counts
  from planar graphs.
\newblock In {\em 2020 IEEE 61st Annual Symposium on Foundations of Computer
  Science (FOCS)}, pages 661--672. IEEE, 2020.

\bibitem[Orc14]{orchardThesis}
Dominic Orchard.
\newblock {\em Programming contextual computations}.
\newblock PhD thesis, 2014.

\bibitem[Pai20]{paine2020}
Tom Paine.
\newblock A pebbling comonad for finite rank and variable logic, and an
  application to the equirank-variable homomorphism preservation theorem.
\newblock In {\em Proceedings of MFPS}, 2020.

\bibitem[RS83]{robertson1983}
Neil Robertson and P.D. Seymour.
\newblock Graph minors. i. excluding a forest.
\newblock {\em Journal of Combinatorial Theory, Series B}, 35(1):39--61, August
  1983.

\bibitem[She71]{shelah1971}
Saharon Shelah.
\newblock Every two elementarily equivalent models have isomorphic ultrapowers.
\newblock {\em Israel Journal of Mathematics}, 10(2):224--233, 1971.

\bibitem[WL68]{weisfeiler1968}
Boris Weisfeiler and Andrei Leman.
\newblock The reduction of a graph to canonical form and the algebra which
  appears therein.
\newblock {\em NTI, Series}, 2(9):12--16, 1968.

\end{thebibliography}

\end{document}